\documentclass[apj]{emulateapj}
\usepackage{apjfonts}
\usepackage{natbib}
\usepackage{graphicx}
\bibpunct{(}{)}{;}{a}{}{,}

\begin{document}
   \title{The Evershed effect observed with 0\farcs 2 angular 
	resolution}
   \author{
   	J.~S\'anchez~Almeida\altaffilmark{1},
  	I.~M\'arquez\altaffilmark{1,2},
 	J.~A.~Bonet\altaffilmark{1},
	I. Dom\'\i nguez Cerde\~na\altaffilmark{1}}
    \altaffiltext{1}{Instituto de Astrof\'\i sica de Canarias, 
              E-38205 La Laguna, Tenerife, Spain}
    \altaffiltext{2}{Departamento de An\'alisis
     	Matem\'atico, Universidad de La Laguna, E-38271 La Laguna, 
     	Tenerife, Spain}
   \email{jos@iac.es,imr@iac.es,jab@iac.es,itahiza@iac.es}
  
\begin{abstract}

We present an analysis of the Evershed effect
observed with 
a resolution of 0\farcs 2.
Using  the new  Swedish 1-m Solar Telescope 
and its Littrow spectrograph,
we scan a significant part of a sunspot penumbra.
Spectra of the non-magnetic line  
Fe~{\sc i}~7090.4~\AA\  allows us to measure Doppler shifts 
without magnetic contamination.  
The observed line profiles are asymmetric. The
Doppler shift depends 
on the part of the 
line used for measuring, indicating that the velocity 
structure of  penumbrae remains unresolved even with 
our 
angular 
resolution.
The observed 
line profiles are properly reproduced 
if two components with velocities between 
zero and several ~km~s$^{-1}$  
co-exist in the resolution 
elements. Using Doppler shifts at fixed line
depths, we find a local 
correlation between upflows and bright structures,
and downflows and dark structures. 
This association is not specific of the outer penumbra but it 
also occurs in the inner penumbra. 
The existence of  such correlation was originally reported 
by Beckers \& Schr\"oter~(1969), and it is suggestive
of energy transport by convection in penumbrae. 

\end{abstract}

%
\maketitle

\section{Introduction}\label{intro}

The spectral lines observed in sunspot penumbrae 
are both asymmetric and shifted in 
wavelength. This 
phenomenon, known as the Evershed effect, 
is commonly attributed to the presence 
of large and highly inclined plasma flows.
Although the effect was discovered a hundred years ago 
\citep{eve09} the true nature of these flows remains 
enigmatic, which hampers a full understanding 
of the penumbral physics.
The observational properties of penumbrae and the 
Evershed effect are well covered in a number of
recent review papers \citep[e.g.,][]{tho92,tho04,sol03},
and we refer to them for an overview.
This introduction is focused on a particular aspect 
of the effect
directly connected to our observational work.

Penumbrae are almost as bright as the quiet Sun. Since 
the transfer of energy by radiation is  
inefficient, either the penumbrae are transparent and 
therefore very shallow, or the energy radiated away by 
penumbrae is transported from below by some sort 
of convective process\footnote{
Here and throughout the text we adopt the loose
definition of convection by \citet{nor00};
"convection is the transport of energy by hot fluid 
moving upwards and cold fluid moving downwards". 
\citep{spr87,sch86b}. 
}
Shallow penumbrae require nearly horizontal magnetic fields,
which are not observed and which leaves us only the second 
option \citep{sol93c}. Therefore, the penumbrae of sunspots,
with their filaments and flows, are likely 
the result of  
convection
taking place in highly inclined 
strong magnetic fields. However,  there is no 
agreement on how the convection operates, i.e., on how 
the hot plasma rises, releases energy, and returns 
to the sub-photospheric layers, and how these motions
occur in a strong magnetic field constraining 
the free movement of the plasma. 
Several possibilities have been put forward
in the literature, and we will mention a number of 
them as example. The penumbrae may contain 
convective rolls, where the magnetic field lines are
rigidly transported by the plasma motions, and which require 
very inclined
magnetic fields \citep{dan61}. Depending on the magnetic field 
inclination, convection sets in  
as convective rolls or traveling
waves \citep{hur00}. 
Siphon flows along magnetic field lines have 
been proposed mainly for explaining the horizontal
component of the Evershed
flow \citep{mey68,tho93,wei04}. However,
such flows transport hot plasmas from 
the sub-photosphere, which become optically thin
and radiate away internal energy. 
Individual  magnetic fluxtubes that raise, cool off, and then 
fall represent another possibility. This process is called
interchange convection by moving magnetic fluxtubes 
\citep{sch91,jah94}, and it has been modeled in 
detail by 
\citet{sch98a,sch98b}. 
A significant part of the convective 
energy transfer is not provided by the interchange motions, 
but by flows along field lines indirectly induced 
during the interchange \citep{sch03f}. 
In the vein of the thin penumbra models \citep{sch86b}, but 
improved to get rid of the problems, \citet{spr06} propose the
intrusions of non-magnetic convective cells.

None of the modes of convection mentioned above
satisfy all the observational constraints and, 
simultaneously, none of the observations seem to
be so deciding as to rule out particular models once and
for all.
The main problem probably
lies in the insufficient spatial 
resolution of our observational description of the 
phenomenon. Convection could be easily identified 
upon detecting a clear correlation between vertical 
velocity and brightness, similar to that 
characteristic of the solar  granulation.  
However, the flows in the penumbra are
known to change direction and speed 
within scales smaller than the photon  
mean-free-path, a property inferred directly from
observations (e.g., from the existence 
of broad-band circular 
polarization, \citealt{san92b}; see also 
\S~\ref{discusion}). The photospheric mean-free-path
length scale, some 150~km,  is comparable with the
angular
resolution of the best images, therefore, even our best 
measurements are expected to average several 
structures with different 
physical properties. 
The observations
always provide ill-defined averages of the 
measured properties,  being each resolution element a
volume of the solar photosphere. The size in the plane 
perpendicular to the line-of-sight (LOS) is set by the
angular  resolution, whereas  the size along the LOS is 
set by the radiative transfer smearing 
\citep[e.g.,][]{san98c}. Unscrambling the averages to 
retrieve properties of the unresolved structure
is both difficult and prone to misinterpretations.           
Researchers have approached the problem assuming 
the existence of unresolved
structures during the data analysis
\citep[e.g.,][]{bum60,gri72,gol74,san92b,sol93b,
wie95,san96,wes01b,
bel04,san04b}. Then the outcome of the
analysis depends on the specific assumptions, a 
difficulty that increases as the resolution 
worsens and the observational 
average becomes coarser. On the contrary, improving 
the angular resolution removes 
part of the 
ambiguities and 
facilitates the interpretation of the
observations. In view of past
experiences, it is difficult to specify an
observational plan  that secures sorting out
the penumbral puzzle.  However, any 
successful plan necessarily requires improving 
the angular
resolution to a point where the 
interpretation of the observations is not
longer ambiguous.

The advent of the Swedish 1-m Solar Telescope
\citep[SST; ][]{sch03c,sch03d} has opened up  
new possibilities since it routinely provides 
an angular
resolution significantly better than
its predecessors \citep[][]{sch02}.
We take advantage of the SST to carry out 
spectroscopic observations of the 
Evershed effect with an exceptional spatial 
resolution ($\simeq$0\farcs 2; \S~\ref{observations}). 
Our study is focused on the local
correlation between brightness and Doppler shift 
originally found by \citet{bec69c} and later 
observed by others \citep[][]{san93b,joh93,sch99b,sch00b}. 
Features brighter than the local mean are associated 
with blueshift, and vice versa. The same correlation 
exist both in the limb-side penumbra and the center-side 
penumbra, a fact invoked by \citet{bec69c} to 
conclude that vertical motions rather than horizontal 
motions are responsible for the correlation. Such
correlation  between vertical velocity and intensity 
is characteristic of the granulation. 
The fact that the same correlation is also present 
in penumbrae suggests a common origin for the 
two phenomena, namely, convection. Despite the 
qualitative agreement, the amplitude of the velocity 
fluctuations, of the order of  100~m~s$^{-1}$,
is insufficient to transport  the energy radiated away 
by penumbrae. Velocities similar 
to those of the non-magnetic granulation are
required independently of the specific mode of convection
responsible for the transport
\citep[see, e.g.,][]{spr87}. 
We revisit the issue with the best resolution available 
at present to (a) confirm  the existence of such correlation,
and (b) see whether the improved resolution
reveals vertical velocities large enough to account 
for the radiative losses of penumbrae.
With this in mind,
we seek independent support for various
radically different observational studies suggesting
the presence of convective motions
in penumbrae.
Filament like structures in velocity maps
are found to join bright and dark knots 
\citep{sch04b,sch05}. 
A complex velocity pattern 
with small-scale upflows and downflows
is able to reproduce the asymmetries of the Stokes 
profiles characteristic of penumbrae 
\citep{san04b}.
The same scenario is qualitatively consistent with
the proper motions observed in a penumbra 
by \citet{mar05}. The proper motions  are 
predominantly radial but 
with a sideways component which arranges 
the plasma forming long narrow filaments co-spatial with
dark features.  Mass conservation arguments indicate 
that the accumulation of plasma in the dark filaments must 
be balanced by downflows of the order of 200~m~s$^{-1}$.

Our study of the Evershed effect begins by 
describing the observations and the data reduction 
(\S~\ref{observations} 
and Appendix~\ref{appb}).
The shape 
of the observed line bisectors is analyzed and
discussed in \S~\ref{bisecsec}. The observed spectra 
show the  correlation between Doppler shift and brightness 
discussed above. Its variation with the azimuth in 
the penumbra allows us to separate vertical
and horizontal velocities (\S~\ref{vert_and_hor}).
Line asymmetries are analyzed 
in \S~\ref{two_gauss_fit} in terms of two spatially
unresolved components. The implications of our work
for understanding the structure of penumbrae 
are discussed  in \S~\ref{discusion},  
where
we also consider consistencies and 
inconsistencies with previous studies. 
Finally, Appendix~\ref{appa} shows that the kind of 
weak correlation between vertical motions and 
brightness found in penumbra is also observed in the
quiet Sun when the spatial  resolution is insufficient 
to resolve individual granules.

%
\section{Observations and data reduction}\label{observations}
\begin{figure*}
\plotone{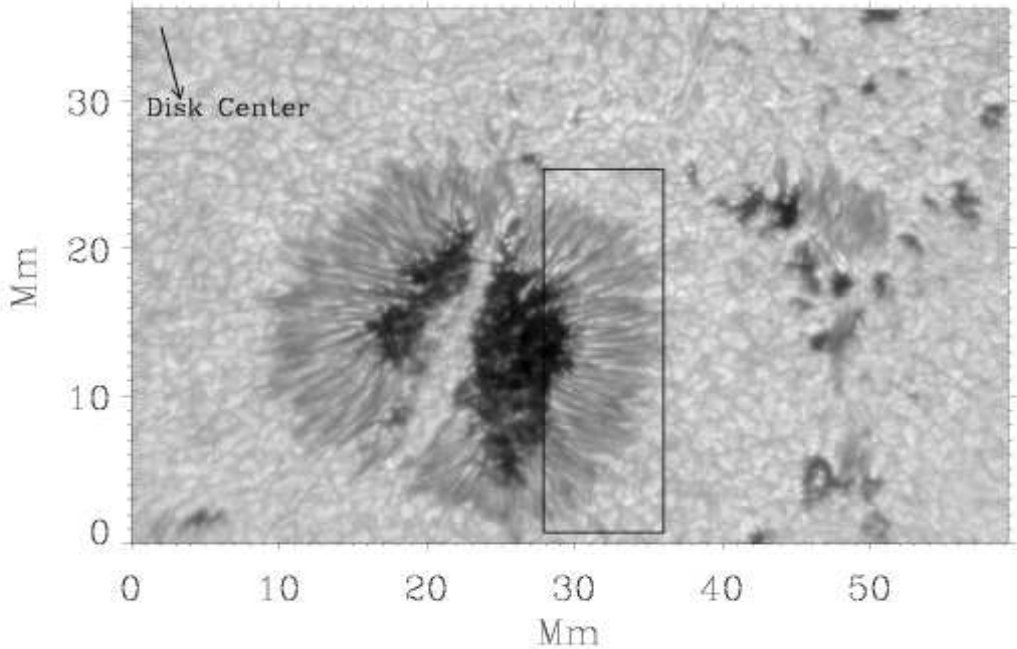}
\caption{Slit-jaw of one of the exposures showing our field
of view (the box). The axes are given in Mm from the
lower left corner, and an arrow points out the direction
of the solar disk center.
The scan was carried out from left to right with the 
slit in the vertical direction.
(The slit was artificially removed from this
slit-jaw image during the flatfielding procedure.)
The image was obtained with a 7~\AA\ wide color filter
centered in 7057~\AA .
} 
\label{slit_jaw}
\end{figure*}

The observations were carried out on October 16, 2004, 
using the SST equipped  with 
AO and a Littrow spectrograph 
\citep[][\footnote{
TRIPPEL spectrograph
; see 
http://www.solarphysics.kva.se/}]
{sch03c,sch03d,kis06}.
We obtained spectrograms of the non-magnetic line 
Fe~{\sc i}~7090.4~\AA\ in the penumbra of the 
decaying circular sunspot NOAA 10682 (heliocentric
angle $\theta\simeq 35^\circ$;  $\cos\theta$=0.82).
The black rectangular box in Figure~\ref{slit_jaw}
indicates the field-of-view (FOV).
We employ a non-magnetic line to avoid contaminating 
the inferred velocity fields with cross-talks 
coming from the penumbral magnetic field. 
The penumbra was scanned from the inner to the outer 
boundaries with steps of 0\farcs 12 perpendicular to 
the slit. 
The slit width was set to 0\farcs 11, which corresponds 
to 29~m\AA\ in the focal plane of the spectrograph and 
sets the spectral resolution of the data.  
Given the pixel size of the camera and 
the effective focal length at the 
focal plane of the spectrograph, the sampling
interval of the spectra is 0\farcs 04 along
the slit and 10.5~m\AA\ across the slit, 
which suffices for the spatial and spectral 
resolution of the data.
We use frame selection triggered by the 
spectrograph camera, providing the four
spectra of highest contrast every 
20.8~s. The 
exposure time was set to 150~ms.
The scan covers an effective FOV of 
10\farcs 5 (across the slit) $\times$  34\farcs 4 
(along the slit), and it required 31~min to 
be completed.
Together with the spectrograms, we gather simultaneous 
slit-jaw and Ca~H 
images (an example of the former
is shown in Fig.~\ref{slit_jaw}). They were used
 to set up the polar coordinate system used  
for the analysis of velocities carried out 
in \S~\ref{vert_and_hor}, which requires identifying  the 
direction of the solar disk center,
as well as assigning a curvature radius and 
a curvature center to the penumbra. 
The data are corrected for dark current and
flatfield using standard procedures. 
In addition,
we correct the spectra for the Modulation Transfer
Function (MTF) characteristic
of an ideal 1-m telescope at the working wavelength.
Since the penumbral structures are predominantly 
elongated across the slit, the deconvolution process
can be reduced to a 1D problem using as MTF 
a radial cut across the 2D axi-symmetric MTF of 
a diffraction limited telescope
\citep[see, e.g., ][]{san98b}.
The deconvolution, including optimum filtering,
is carried out in the Fourier domain following the
procedure explained in detail by \citet{sob93}.

\begin{figure*}
\includegraphics[angle=90.,width=0.95\textwidth]{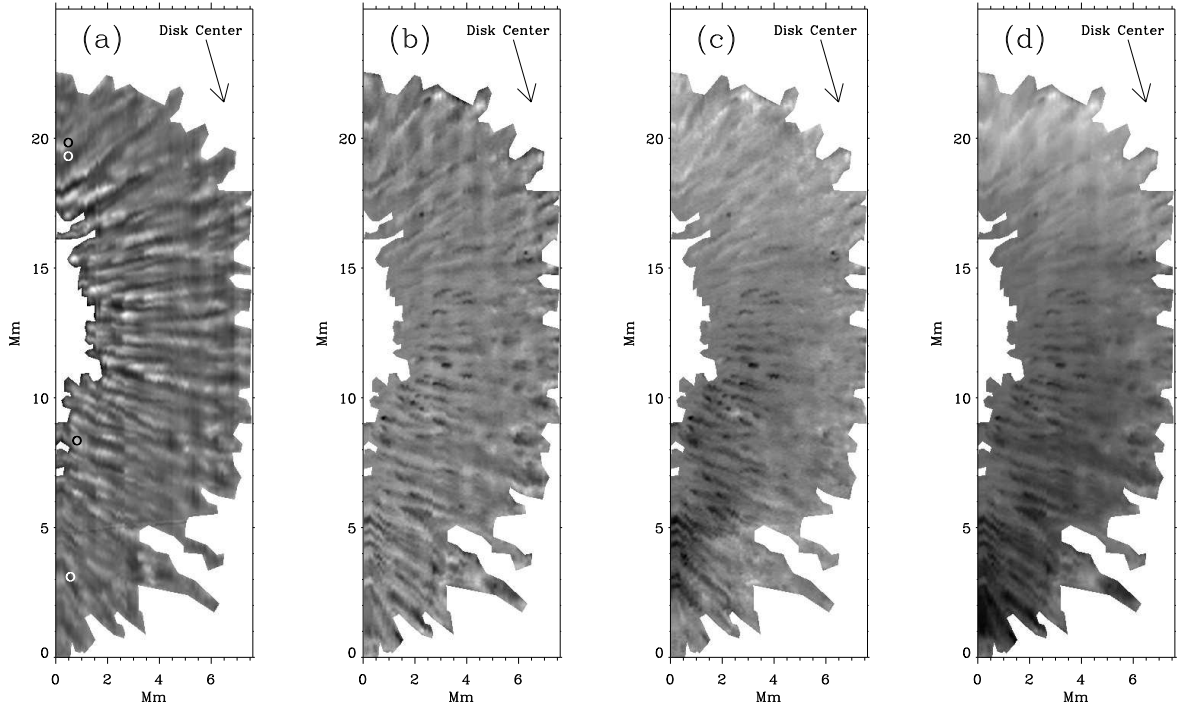}
\caption{
(a) Map of continuum intensity. (b) Doppler shift of the 
bisector at 80\%  of the line 
depth. Running means have been subtracted in these
two maps to bring out only local variations.
(c) Difference between 
the Doppler shifts of the bisectors  at 80\% and 20\%
without running mean subtracted. Images
(b) and (c) share the same scale of grays which
goes from -2~km~s$^{-1}$(black) to +2~km~s$^{-1}$(white).
(d) Original map of the Doppler shift without the
running mean subtracted and represented 
with a scale spanning
from -4~km~s$^{-1}$(black) to +4~km~s$^{-1}$(white).
According to our convention, redshifts correspond
to positive Doppler shifts. The centers 
of the symbols in
(a) indicate the positions of the representative
line profiles shown in Figure~\ref{bisectors}. 
}
\label{my_1_fig}
\end{figure*}
\begin{figure}
\includegraphics[width=9cm]{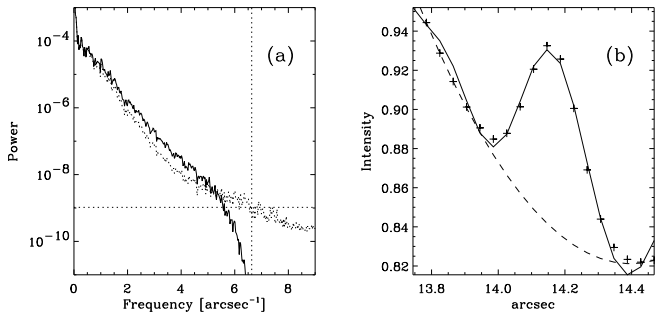}
\caption{
(a) Power spectrum for the intensity variations
along the slit. The dotted line shows the   
average among all our penumbral spectra before restoration.
The solid line represents the same average after restoration.
The horizontal dotted line corresponds to the level of power
at the cutoff imposed by the
telescope diameter and, therefore, it indicates the level
of noise. There is power in the original unprocessed image
even beyond 5~arcsec$^{-1}$, corresponding to a 
scale of 0\farcs 2. The restoration includes a noise filter
that damps down 
the spectrum starting at this threshold. 
(b)  The solid line shows observations of the 
intensity variation across a representative narrow 
structure in one of the spectra. The symbols correspond to
a least squares fit of this profile with a Gaussian plus
a parabolic background, the latter represented by the 
dashed line. The Gaussian has 0\farcs 19 FWHM, indicating 
that features of this width can be resolved in our spectra.
}
\label{cutoff}
\end{figure}

From the best spectrogram per slit position,
we construct the 2D continuum image of the 
FOV shown in Figure~\ref{my_1_fig}a.
Each column corresponds to one 
slit position so that the full set provides an 
image. Residual image motion in the direction parallel 
to the slit has been removed from the image
by cross-correlation techniques. We shift the different 
columns so that the cross-correlation between 
adjacent columns is maximum. 
The same small re-arrange of the columns inferred from the 
continuum intensity is applied to all the maps displayed 
and discussed in the paper. 
This manipulation was 
carried out mainly for aesthetic reasons, 
since the residual image motion has  very limited 
impact on the results of our analysis, which is based 
on correlations between  parameters obtained in 
the same pixels. The angular resolution of our 
spectra turns out to be very close to 0\farcs 2, as
inferred from the cutoff frequency of the power spectra 
of continuum fluctuations in the spatial direction of the 
spectrograms. Figure~\ref{cutoff}a shows 
the power spectrum for the variations
of intensity along the slit. We average the
power spectrum of all the columns forming the
continuum image in Figure~\ref{my_1_fig}a. Both the original
spectra and the MTF reconstructed spectra show signals above the
noise at  5~arcsec$^{-1}$, corresponding to a 
scale of 0\farcs 2. The same conclusion can be 
drawn from Gaussian fits to the narrowest continuum 
features in the spectra. Figure~\ref{cutoff}b shows 
one of these features reproduced with a  Gaussian 
having a FWHM of 0\farcs 19. 

The work aims at analyzing absolute Doppler shifts,
which requires a careful calibration of the 
wavelength scale. First, the relative wavelength
scale along the spectrograph slit was set using
spectrograms taken while the telescope pointing was
moving across the solar disk.  
The spectral line 
must have the same wavelength in these flatfield images, 
an information used to bring to a common scale
all the spectra taken in each position of the
scan. As we argue below, the spectrograph was stable enough 
during the scan to allow a direct comparison of the wavelengths 
taken at different positions, which provides a consistent
relative 
wavelength scale for the whole scan. Therefore, one can set 
a common  {\em absolute} scale by knowing the 
absolute wavelength of only one spectrum. We set such scale assigning 
zero velocity to the mean line shift obtained from
the minimum of Fe~{\sc i}~7090.4~\AA\ in the
darkest parts of the umbra. According to \citet{bec77},
such absolute velocity is
uncertain by less  than 100~m~s$^{-1}$.
Umbral oscillations may have an influence on 
the velocity scale thus obtained, but
their amplitudes are again smaller than 
100~m~s$^{-1}$ \citep[e.g.,][]{lit92}.
The effect of the spatial stray-light is analyzed 
in Appendix~\ref{appb}, and it turns out to be
negligible.
We can assure the stability of the spectrograph during
the scan because of the position the  H$_2$O 
telluric line at 7094.05 \AA . 
The wavelength of the minimum of this line in the absolute 
scale used for  Fe~{\sc i}~7090.4~\AA\ is constant with a rms 
fluctuation of only  90~m\,s$^{-1}$. 
Most of these fluctuations result from uncertainties
in determining the minima of the weak telluric line
(equivalent width of only 5 m\AA). In short, our
absolute wavelength scale provides velocities
with an accuracy of the order of 100~m~s$^{-1}$.

\section{Bisectors and unresolved velocities}\label{bisecsec}

\begin{figure}
\plotone{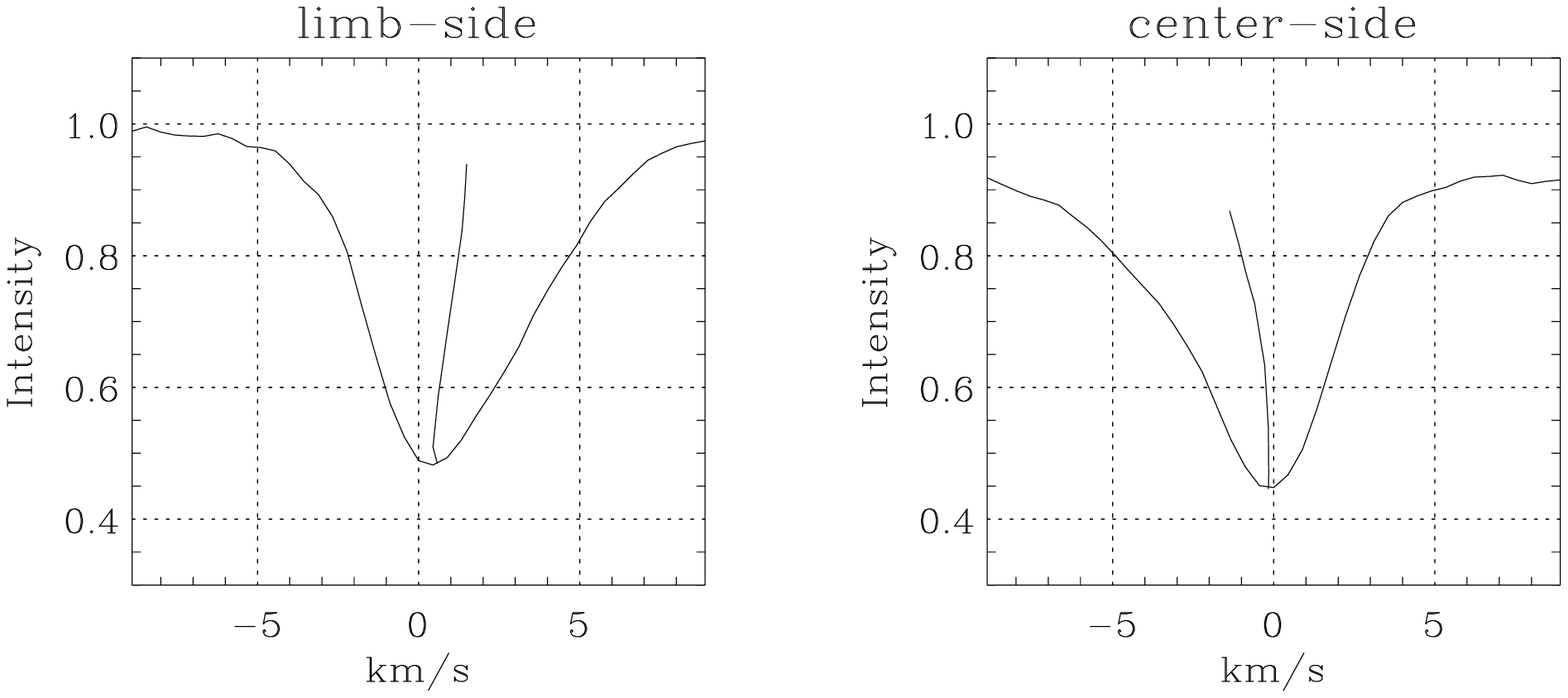}
\plotone{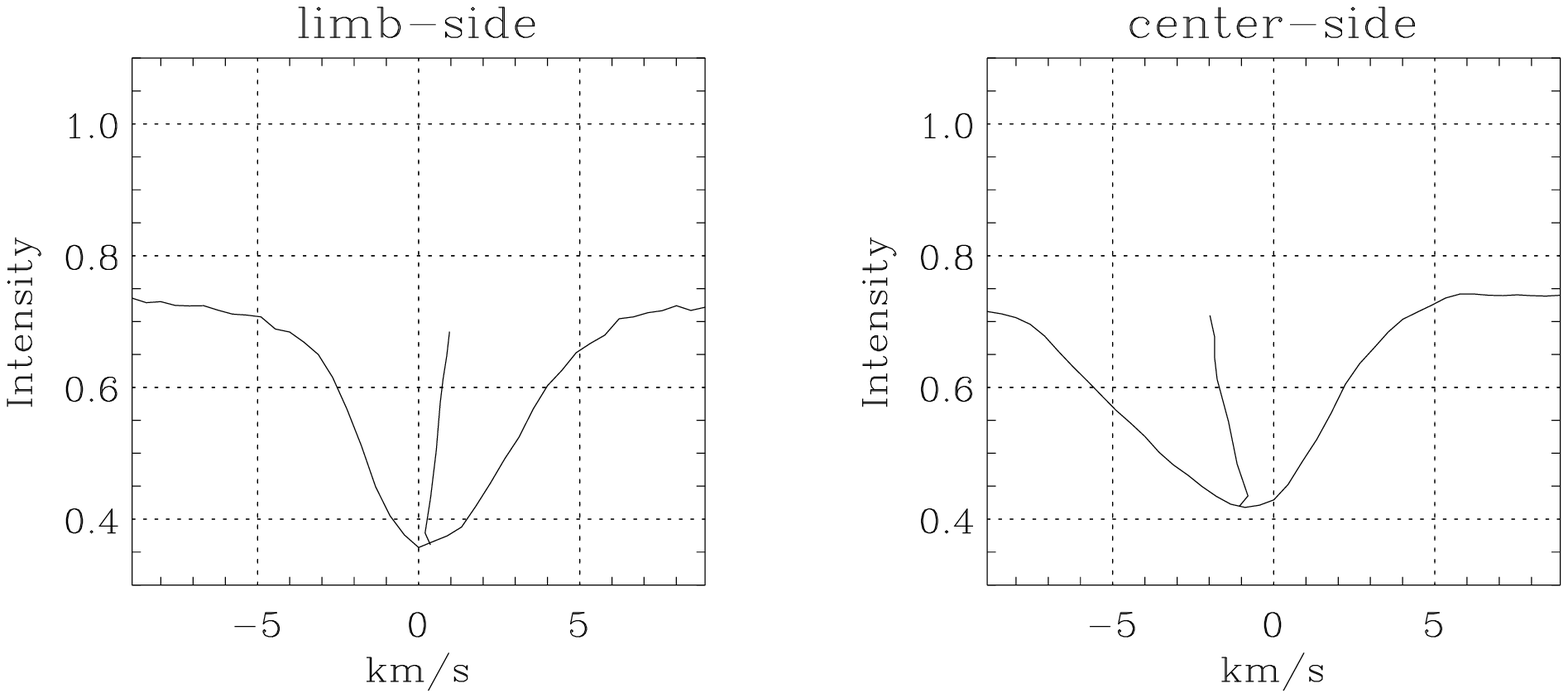}
\caption{Representative line profiles showing large
asymmetries and the corresponding bisectors.
Top: profiles of bright features. Bottom: profiles
of dark features. Left: profiles on the limb-side penumbra.
Right: profiles on the center-side penumbra.
The actual positions are indicated by the symbols
in Figure~\ref{my_1_fig}a.
Wavelengths are 
in km~s$^{-1}$ in our
absolute velocity scale,
whereas the intensities are 
given in units of 
the mean quiet Sun 
continuum intensity.
}
\label{bisectors}
\end{figure}

Spectral lines formed in atmospheres of constant
velocity are symmetric and shifted as a whole,
therefore, one should get the same Doppler shift
independently of the part of the line used for 
measuring. Since  this is not the case of 
our penumbral spectra, we conclude that even
with 0\farcs 2 angular resolution a significant
part of the penumbral velocity structure remains 
unresolved. 
We bring up this fact by computing the
bisectors of the intensity profiles.  
The bisector of an intensity profile is the line joining 
the mid points of the profile at each intensity level.
(The intensity level is usually quantified using a 
percentage, with 0\% representing the line core
and 100\% the continuum.)
Figure~\ref{bisectors} shows a few representative
examples of observed line profiles and bisectors.
The bisector must be a vertical straight line if the atmosphere
has a constant velocity, however, the bisectors of 
Fe~{\sc i}~7090.4~\AA\ are curved and inclined. They show 
different velocities 
at different line depths, indicating that the velocity structure
is finner than our resolution.
The lack of enough resolution
is not an exception attributable to a few pixels. 
Figure.~\ref{my_1_fig}c
shows a map of the velocity difference between the bisector at
line wings (80\%) and the bisector at line core (20\%).
The map of differences 
follows a pattern with
the characteristic filamentary appearance 
of the penumbra. 
The typical values differ
from zero, 
and values as large as one 
km~s$^{-1}$ are not unusual.
The dispersion of velocities responsible
for these bisectors  must be large, a
conclusion that we try to quantify 
with the two-Gaussian fits presented in 
\S~\ref{two_gauss_fit}. 

As it is pointed out
in \S~\ref{intro},
our resolution element is actually 
a volume, with the size in the plane perpendicular
to the LOS set by the angular resolution, 
and the size along the LOS set by radiative transfer
smearing. Consequently, observing slanted bisectors do not 
allow us to know if the gradients of velocity
occur along or across the LOS. The
two of them probably coexist (see \S~\ref{discusion}).
\section{Relationship Between 
Continuum Intensity and Velocity}\label{vert_and_hor}

In agreement with \citet{bec69c} and
others \citep[e.g.][ see \S~\ref{intro}]{san93b,joh93,sch00b},
we find a local correlation 
between continuum intensity $I$ and Doppler shift $U_D$.
The purpose of the section is twofold. First, it
describes the observed correlation.  Second,
horizontal and  vertical motions are 
separated to show that a significant part of the 
correlation is due to vertical motions.  

\begin{figure*}
\plotone{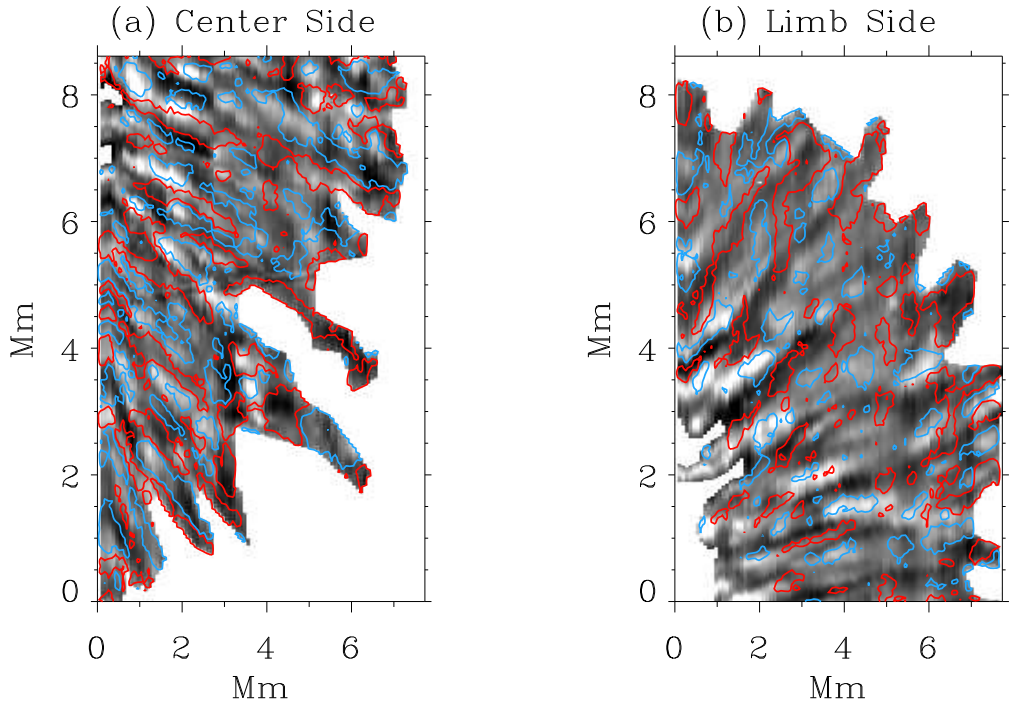}
\caption{
Continuum images of a portion of the
center-side penumbra (a) and the limb-side penumbra (b)
overlaid with contours of the velocity field
inferred from bisector shifts at line wings
(Fig.~\ref{my_1_fig}).
Both the intensities and
the velocities represent local variations; the large scale 
variation has been removed by subtracting a smoothed 
version of the original maps. Blue and red contours correspond to
blueshifts and redshifts, respectively  
($\pm$250~m~s$^{-1}$).
The velocity contours show a clear filamentary structure. 
Note that blueshifts tend to be associated with bright 
features, and vice-versa. It is not a one-to-one correlation,
though. Such association is more
pronounced in the limb-side penumbra, but it is also 
present in the center-side penumbra, indicating that it 
must be produced by vertical rather than horizontal
velocities. 
}
\label{ita_like}
\end{figure*}

The local variations are obtained 
by subtracting from the original maps a running mean.
Figure~\ref{my_1_fig}b has been computed
from the original Figure~\ref{my_1_fig}d 
using a running box 1\farcs 6 wide.
The same box is used for the intensity in
Figure~\ref{my_1_fig}a.
We also tried with boxes half and twice this
value to conclude that the actual width is not 
critical (see below).  
The correlation can be observed directly 
from inspection of the velocity and
intensity images, e.g., Figure~\ref{ita_like}. 
It shows how local blueshifts are preferentially associated with
local bright features. The correlation is more clear
in the limb-side penumbra (Fig.~\ref{ita_like}b), but it
is also present with the same sign
in the center-side penumbra (Fig.~\ref{ita_like}a).
The relationship can be quantified as,
\begin{equation}
U_D-\langle U_D\rangle \simeq m~(I-\langle I\rangle)  + k, 
\label{myeq} 
\end{equation}
with the angle brackets denoting local 
running mean averages.
In this equation and throughout the paper the 
intensity $I$ is referred to the quiet Sun continuum intensity and 
therefore is a dimensionless parameter,
whereas the sign of $U_D$ is chosen to be 
positive for redshifts.

\begin{figure}
\plotone{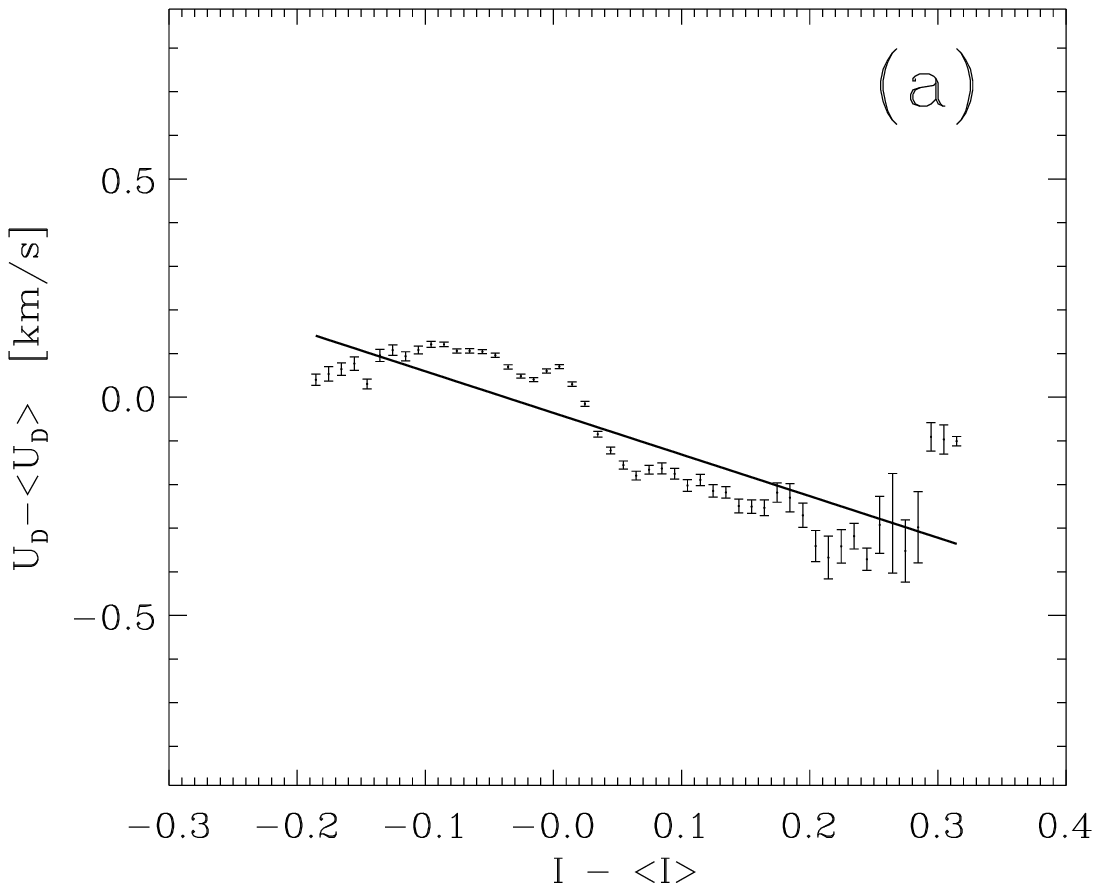}
\plotone{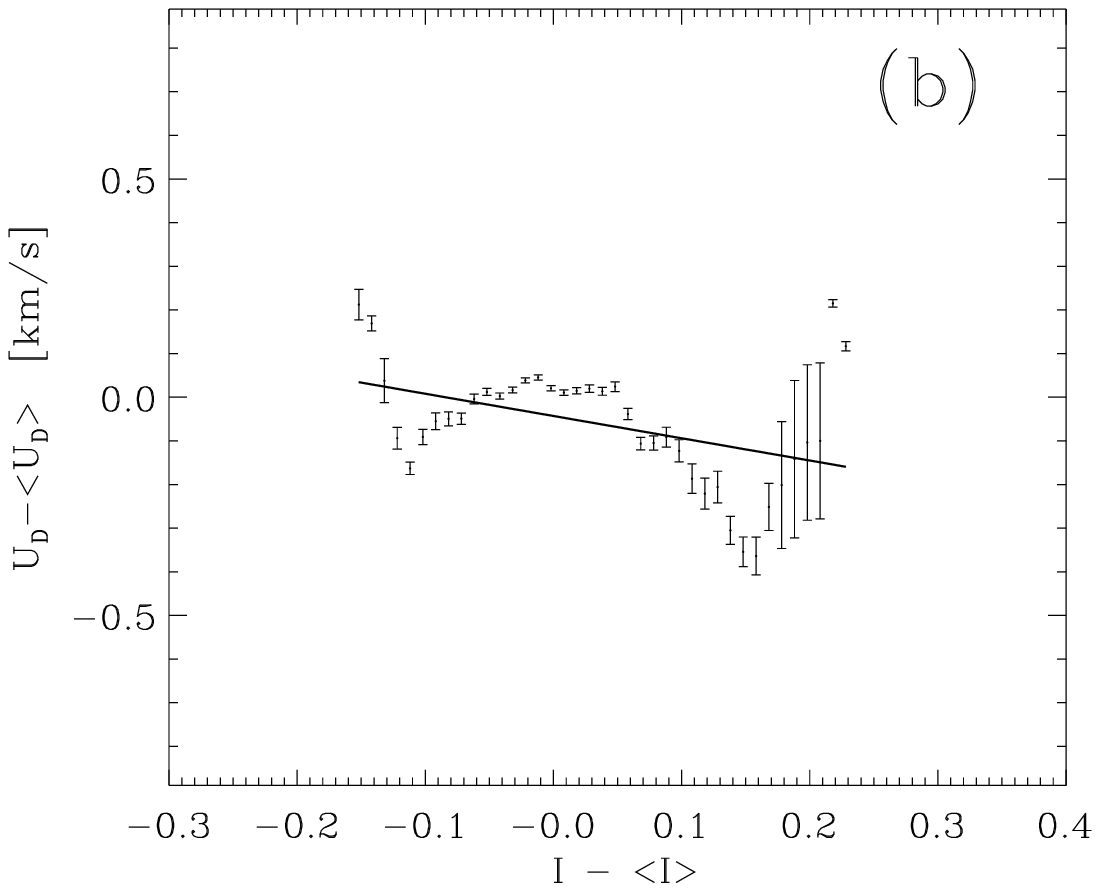}
\caption{Doppler velocity versus continuum 
intensity in sectors at the limb-side penumbra (a) 
and the center-side penumbra (b).
The symbols and the solid lines correspond to observations 
and linear fits, respectively.
The error bars represent standard deviations of the mean
Doppler shift considering all the points with a given 
intensity $I-\langle I\rangle$.
Velocities are measured from bisectors 
at line wings (Fig.~\ref{my_1_fig}b).
}  
\label{my_2_fig}
\end{figure}

Figure~\ref{my_2_fig} shows the observed variation
of the Doppler shift versus continuum intensity for 
two sections of our FOV, one in the limb-side penumbra (a)
and the other in the center-side penumbra (b).
We represent the mean Doppler shift considering intensity
bins of $0.01$. The slope $m$ varies within the sunspot,
from the center-side penumbra to the limb-side penumbra.
However, it maintains a negative sign implying that
bright features are associated with blueshifts 
(Fig.~\ref{my_2_fig}). 
The fact that $m$ keeps the sign implies
that the correlation is produced  by vertical motions, 
since the line-of-sight component of the
horizontal radial velocities changes sign from the center-side
penumbra to the limb-side penumbra \citep{bec69c}.
\begin{figure}
\plotone{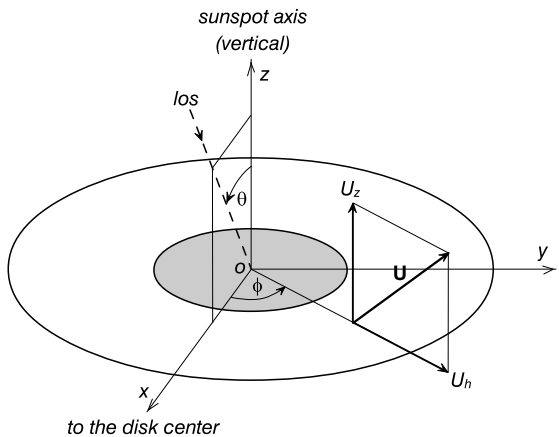}
\caption{Schematic with a sunspot and the 
reference system used to define angles and 
velocities.
The line-of-sight (LOS) corresponding to the
heliocentric angle of observation $\theta$ is shown 
as a dashed line. The velocity {\bf U} of a point
on the penumbra  with an azimuth $\phi$ 
only has a vertical component $U_z$ and a radial horizontal 
component $U_h$.
The sunspot axis coincides with the solar vertical
direction, and the line
joining the sunspot axis and the solar disk center
sets the origin of $\phi$.
}
\label{my_3_fig}
\end{figure}

\begin{figure}
\includegraphics[width=5cm]{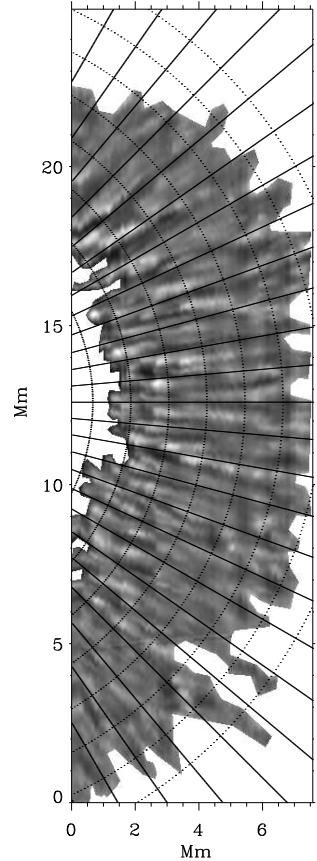}
\caption{Image of the FOV with the grid 
used to assign to the spectra polar coordinates centered on the 
sunspot umbra.}
\label{maparhophi}
\end{figure}
\begin{figure}
\plotone{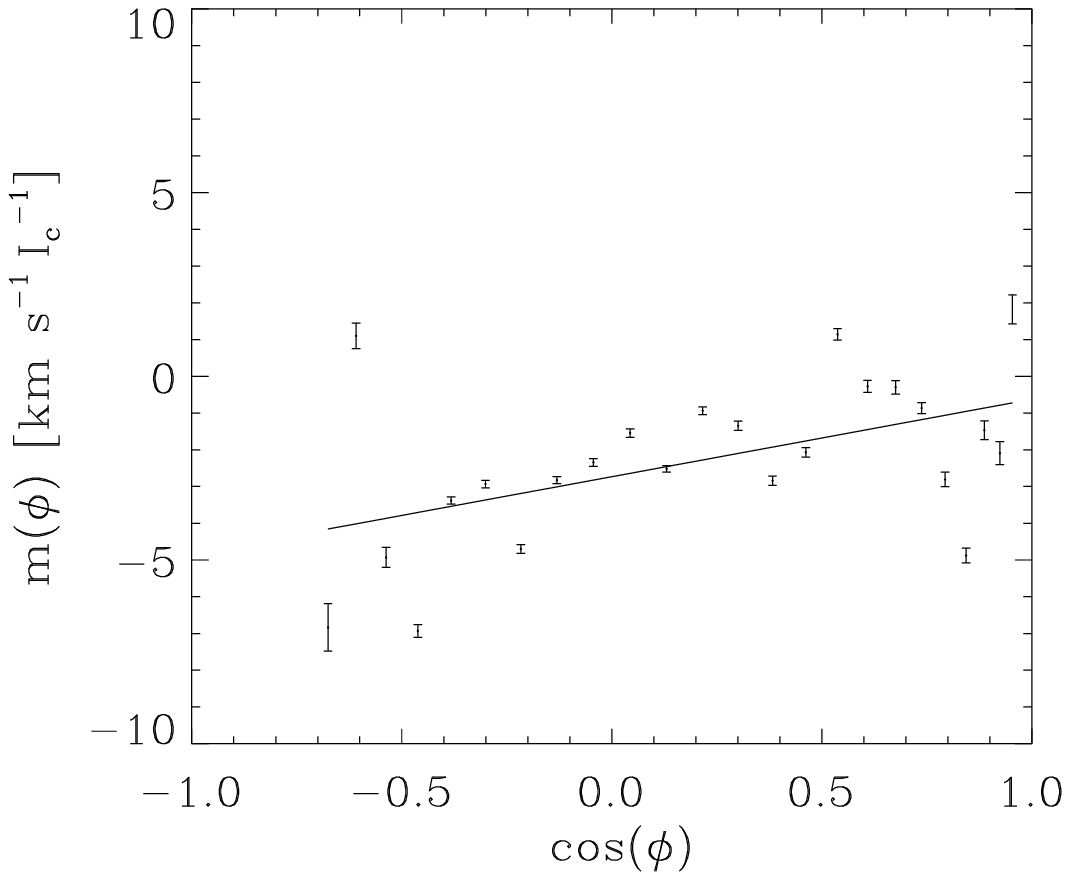}
\caption{
Slope of the relationship between intensity and
Doppler shift $m(\phi)$ versus the cosine of the
azimuthal angle within
the sunspot $\phi$.
The error bars are formal error bars obtained from the fit
used to derive $m(\phi)$.
}
\label{my_3_figb}
\end{figure}
This qualitative argument can be quantified adopting 
the following model for the relationship between 
velocities and intensities,
\begin{eqnarray}
U_z=&m_z\, (I-\langle I\rangle) + k_z,\cr
U_h=& m_h\, (I-\langle I\rangle) + k_h,
\label{def1}
\end{eqnarray}
where $m_z$, $k_z$, $m_h$ and $k_h$ are constants, and
$U_z$  and $U_h$ represent the vertical and horizontal 
velocities in the absolute reference 
system portrayed in  Figure~\ref{my_3_fig}. 
($U_z > 0$ for upward velocities and $U_h > 0$ for
outward velocities.)
If the
motions are predominately radial so that the
azimuthal component can be neglected,
\begin{equation}
U_D=-(U_h\,\sin\theta\,\cos\phi+ U_z\, \cos\theta),
\end{equation}
with the symbol $\theta$ representing the heliocentric 
angle of observation \citep{mal64,sch67}.
This model leads to equation~(\ref{myeq}) with the slope $m$,
\begin{equation}
m(\phi)= -(m_h\,\sin\theta\,\cos\phi+ m_z\, \cos\theta),
\label{my2eq}
\end{equation}
and with the running mean average Doppler shift,
\begin{equation}
\langle U_D\rangle= -(k_h\,\sin\theta\,\cos\phi+ k_z\, \cos\theta).
	\label{avdoppler}
\end{equation}
The average Doppler shifts and slopes depend on the azimuth $\phi$. 
In order to apply the model to the observed velocities, we construct 
a polar coordinate system that assigns azimuths and radial distances 
to each position of the FOV.  
The center of curvature of the observed penumbra was 
estimated visually by trial-and-error so that the grid 
of polar coordinates provides 
radii parallel to the observed penumbral filaments.
The best compromise is shown in Figure~\ref{maparhophi}.
The direction of the solar disk center sets the origin
of the azimuths and it was obtained by
comparison of the slit-jaw images with quasi-simultaneous
full disk MDI continuum images.
Figure~\ref{my_3_figb} shows $m(\phi)$
obtained within the
sectors of approximately constant 
azimuth in Figure ~\ref{maparhophi}
when the Doppler shifts in Figure~\ref{my_1_fig}b are
used (bisector shift at line wings). 
Figures \ref{my_2_fig}a and \ref{my_2_fig}b 
illustrate the kind of fit leading to $m(\phi)$.
Using the model in equation~(\ref{my2eq}), a linear
fit $m(\phi)$ versus $\cos\phi$ renders,
\begin{eqnarray}
m_z\simeq& ~(3.3\pm 0.4)~{\rm km\,s}^{-1},\cr
m_h\simeq& (-3.7\pm 0.8)~{\rm km\,s}^{-1}.
\label{resul1}
\end{eqnarray}
The error bars are 1-sigma formal errors provided by the
least-squares routine, and assume that the deviations
of the data from the fitted line are due to random noise
(see Fig.~\ref{my_3_figb}). 
The sign $m_h < 0 $ indicates that horizontal velocities 
are enhanced in dark features, whereas $m_z > 0 $
implies that upflows occur in bright features.
The fact that $m_h < 0$ and $m_z > 0$ is a firm result, 
as it is indicated by the error bars.
We also checked that bisectors at different line depths
can be used to compute velocities without 
modifying the resulting signs.
The use of bisectors closer to the line core only
reduces the amplitude of the velocities.
We also repeated the calculations using only 
the inner half of the penumbra, i.e., 
the half closest to the umbra.
The signs of $m_z$ and $m_h$ are not modified. 
In addition, we carried out the whole analysis 
doubling and halving the width of the box used 
to compute the running mean. 
Again the signs of $m_z$ and $m_h$ remain. 
Decreasing the width reduces the amplitude 
of the correlation whereas increasing the width
has almost no effect on $m_z$ and $m_h$.

\begin{figure}
%
\plotone{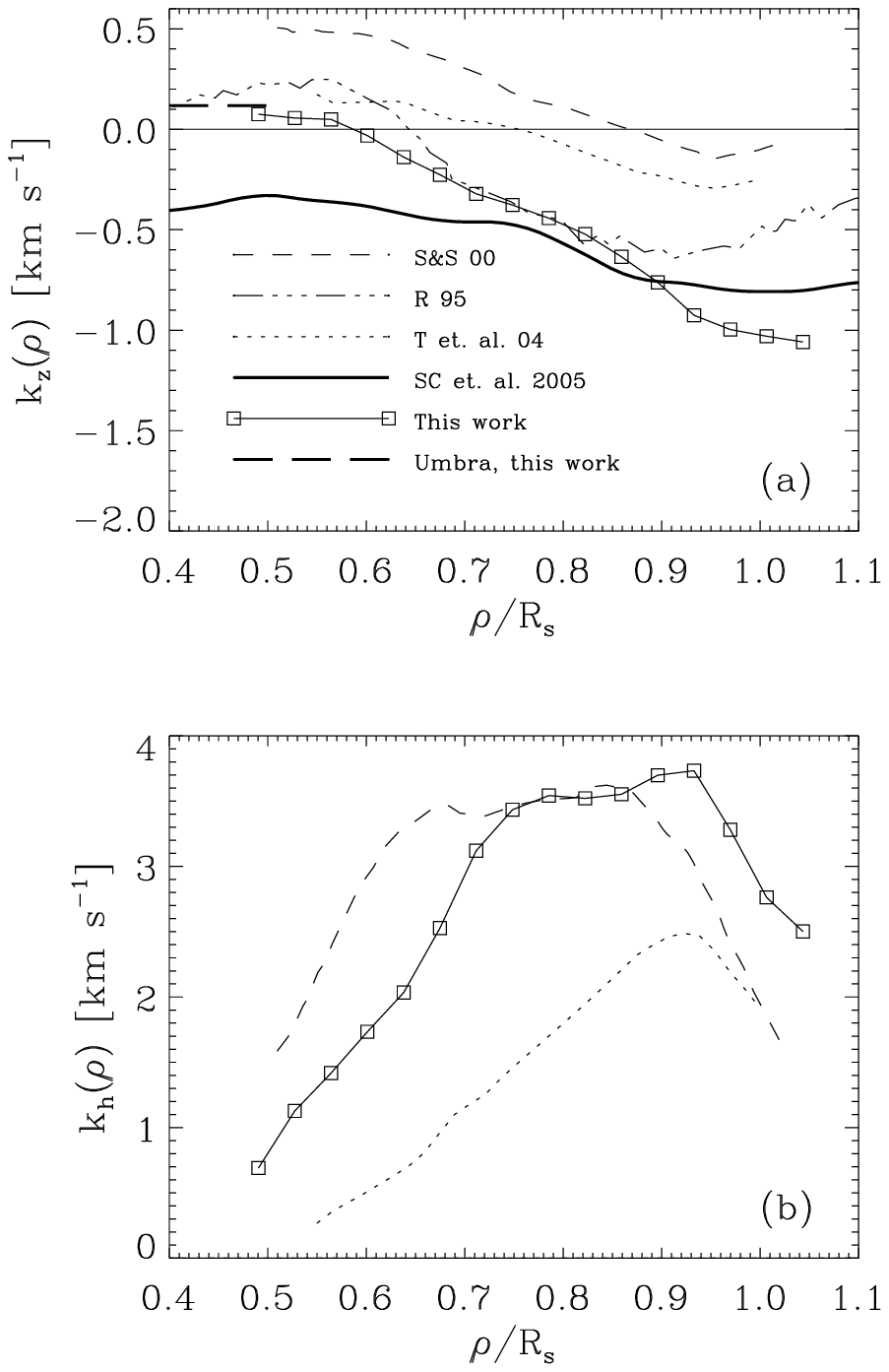}
\caption{(a) Variation of $k_z$
with the distance to the sunspot center (the solid
line with symbols). Note the tendency for $k_z$
to be negative. 
It is not very different from other Doppler shifts
reported in the literature and also included
in the plot. 
The symbols in the inset stand for 
\cite{sch00}~(S\&S 00),
\cite{rim95}~(R 95),
\cite{tri04}~(T et al 04),
and
\cite{san05b}~(SC et al 05). 
(b) Variation of $k_h$ with the distance to the sunspot
center. 
All radial distances are referred to the sunspot radius
$R_s$~($\simeq$14~Mm). 
}
\label{kappas}
\end{figure}

With the same approach leading to $m_z$ and $m_h$, one
can use the local average Doppler shifts 
$\langle U_D\rangle$
to infer the constants $k_z$ and $k_h$ 
(see equation~[\ref{avdoppler}]). It renders,
\begin{eqnarray} 
k_z=&(-0.48\pm 0.05)~{\rm km\, s}^{-1}\cr
k_h=&(3.30\pm 0.15)~{\rm km\, s}^{-1},
\end{eqnarray}
where the error bars are also formal errors from the
linear least-squares fit.
The finding of downward velocities
in penumbrae
($k_z< 0$) is not new. Downward 
velocities have been measured by
many observers starting with the work by 
\citeauthor{rim95}(\citeyear{rim95}; 
see also \citealt{ser61}). 
These downward motions tend to be localized 
in the external parts of the penumbra,
but the actual distribution depends
on the specific measurement.
Figure~\ref{kappas}a shows several among the 
published values
\citep{rim95,sch00,tri04,san05b},
which are presented as vertical velocity
versus distance to the sunspot center.
In order to compare our result with them, we 
consider that $k_z$ and  $k_h$ in equation~(\ref{def1})
depend on the radial 
distance to the sunspot center $\rho$.
Our penumbra is divided in bands of constant radial 
distance (Fig.~\ref{maparhophi}).  Then  equation~(\ref{avdoppler}) was 
fitted to the  azimuthal variation of  the 
running 
mean Doppler 
shift in each one of these bands, which renders  
$k_z(\rho)$ and $k_h(\rho)$. The results have been 
represented in Figure~\ref{kappas}; the solid lines with
symbols, 
with the symbols indicating the center of the band
used to carry out the fits.
We find mild upflows in the inner penumbra and 
downflows in the outer penumbra. These features are 
common to almost all the other measurements portrayed in 
Figure~\ref{kappas}a.
The mean horizontal velocities reach 
values of up to 3.7~km\,s$^{-1}$ not far from the external 
penumbral border, in agreement with values reported in the
literature (Fig.~\ref{kappas}b).
Our curves remain within the scatter of the published 
values, a fact pointed out to
argue that no obvious error 
burdens the validity of our velocity scale.
The quantitative differences between the curves 
represented in Figure~\ref{kappas} can be 
attributed to many factors biasing this type
of observation, e.g., variations from sunspot to
sunspot, different systematic errors in the absolute 
velocity scales, the use of different spectral lines
and procedures to infer velocities, the different ways 
of estimating sunspot radii, etc. 
Although the vertical velocities that we find
are somehow larger than previous estimates 
(Fig.~\ref{kappas}a), the values are close to those
reported by \citet{rim95}
and \citet{san05b}, which correspond to sunspots observed
at the very solar disk center, so that the Doppler shift directly
yields vertical velocities without  geometrical 
transformations. In the particular case of \citet{san05b}, 
they carry out a 1D inversion of the 
spectra to interpret the observed line asymmetries. 
Consequently, they obtain a velocity for each 
optical depth along the LOS. We represent 
the velocity at the bottom of the atmosphere (continuum 
optical depth equals 0.3),
where the line-wings are formed. 

We would like to stress that interpreting  the average 
vertical velocities shown in  Figure~\ref{kappas}a 
is not trivial.  They have been inferred as systematic 
Doppler shifts of the observed spectral lines
(the part of the Doppler shift independent 
of $\cos\phi$; see equation~[\ref{avdoppler}]). 
According to \cite{san04b}, these net 
Doppler
shifts may not 
necessarily represent net vertical plasma motions. 
If the penumbrae have plasma moving up
and down, and if the thermodynamic conditions of these
upflows and downflows are not identical, then there 
will be a net displacement of the average spectrum 
even though there is no net plasma motion. 
This lack of cancellation of the net signal is 
responsible for the  well known convective blueshift of 
the spectral lines observed in the quiet Sun \citep[e.g.][]{dra81}.

\section{Two-Gaussian fit}\label{two_gauss_fit}

\begin{figure}
\plotone{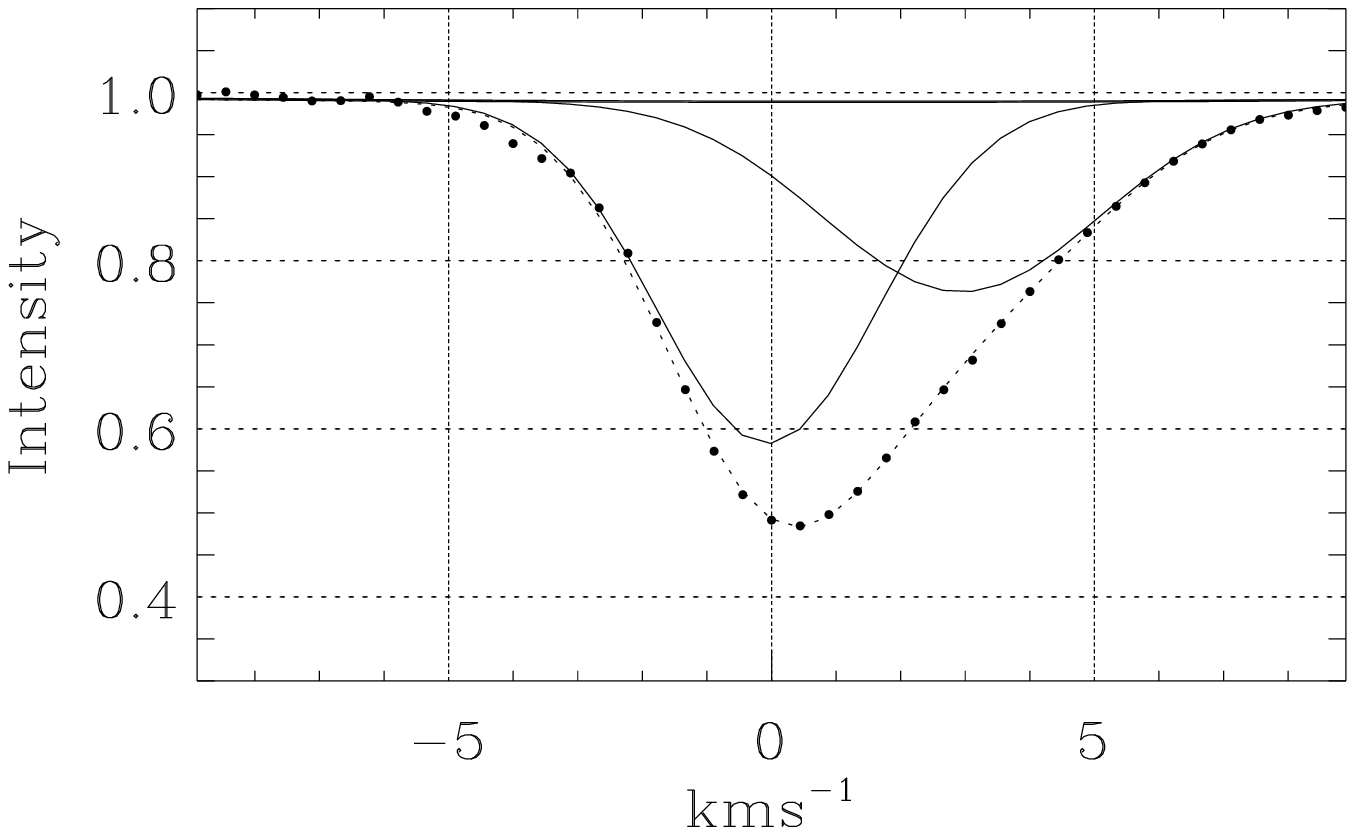}
\plotone{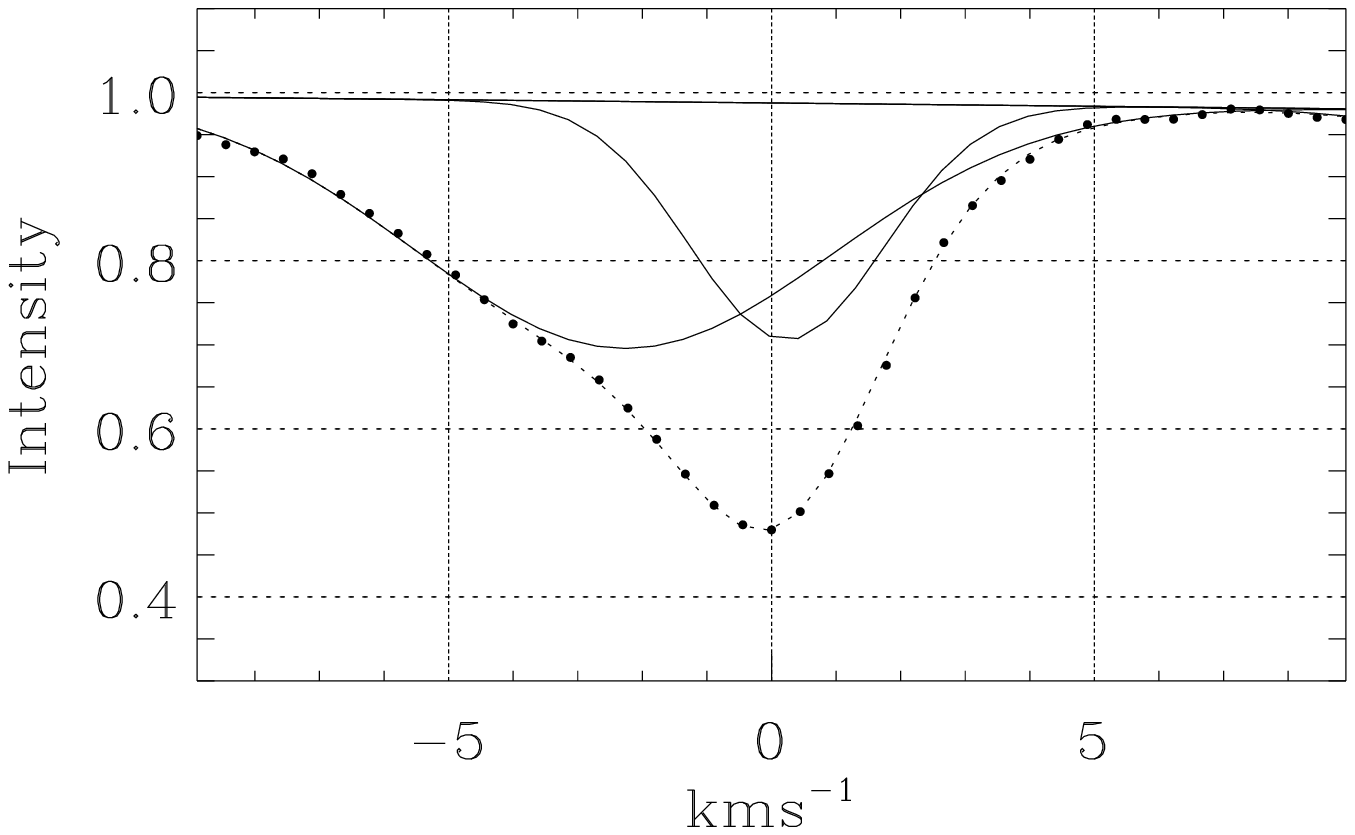}
\plotone{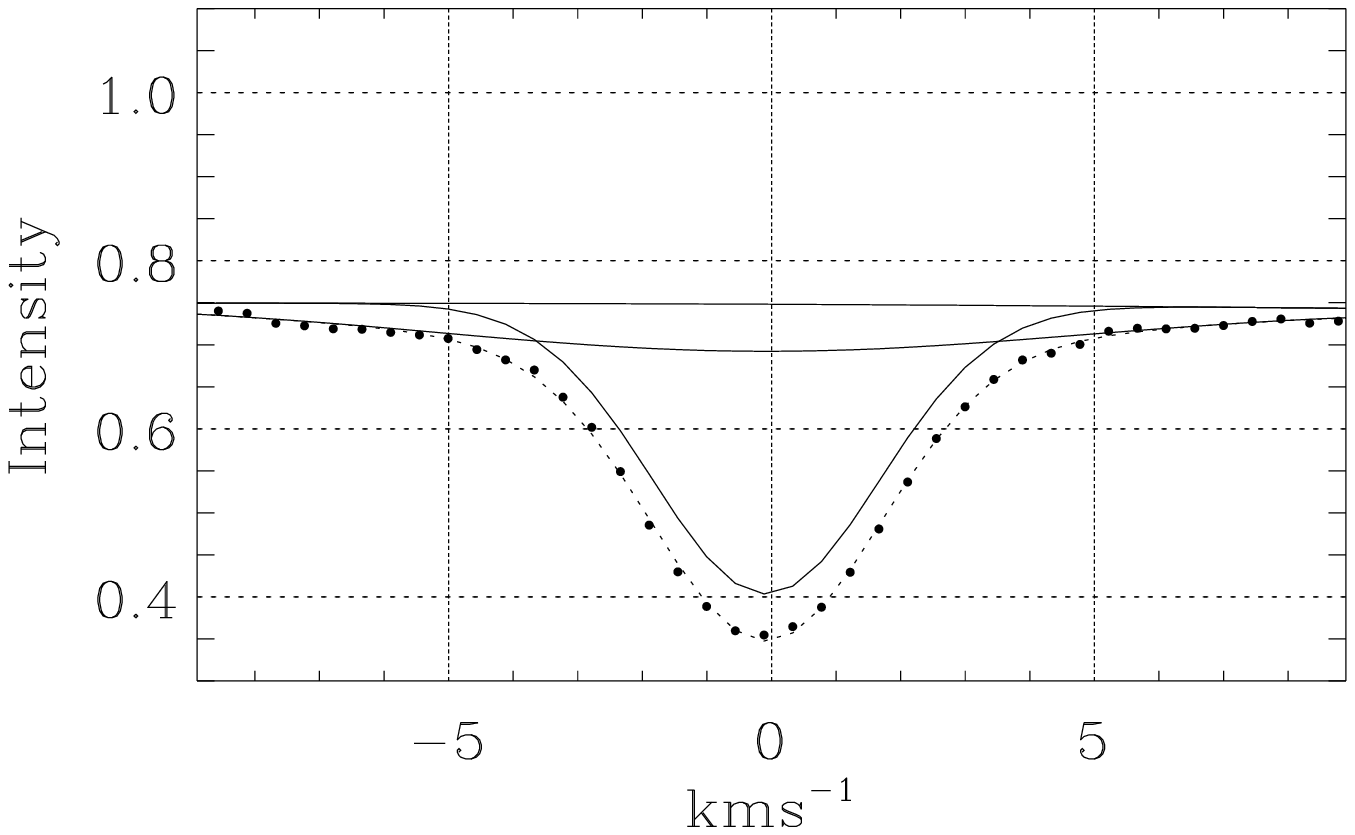}
\caption{
Examples of two-Gaussian fits.
Top: limb-side profile. 
Middle: center-side profile.
Bottom: umbra-penumbra border profile. 
The symbols represent the observations whereas
the dotted line shows the fits. The solid lines 
correspond to the two Gaussians used in the fit. 
	The continuum has been added to the individual
    	Gaussians to facilitate comparison.
}
\label{gauss_fit}
\end{figure}

We also studied whether the observed line
asymmetries can be reproduced by means of two
spatially unresolved velocity components. 
\citet[][]{bum60} puts forward the idea,
and it has been pursued by many others ever since
\citep[e.g.,][]{mal64,ste71,wie95,ich88}.
We fit the line profiles $S(\lambda)$
with two Gaussian functions plus a continuum,  
\begin{eqnarray}
S(\lambda)=&S_c(\lambda)
-d_1\,\exp\big\{-{1\over 2}[(\lambda-\lambda_1)/\sigma_1]^2\big\}\label{eq2g}\\
&-d_2\,\exp\big\{-{1\over 2}[(\lambda-\lambda_2)/\sigma_2]^2\big\}\nonumber,
\end{eqnarray}
with the continuum given by,
\begin{equation}
S_c(\lambda)=a_0+\lambda\ a_1+\lambda^2\ a_2,
\end{equation}
and symbols $\lambda_i, \sigma_i$ and $d_i$ representing,
respectively, 
the central wavelength, the width, and the depth of the two
Gaussians ($i=1,2$).
The fits are in general very good,
like the examples shown in Figure \ref{gauss_fit}.
Note how the symbols used to represent the observations
cannot be distinguished from the dotted line representing
the fits, meaning that the two Gaussians are able to 
reproduce the asymmetries of the observed line 
profiles. 

\begin{figure}
\plotone{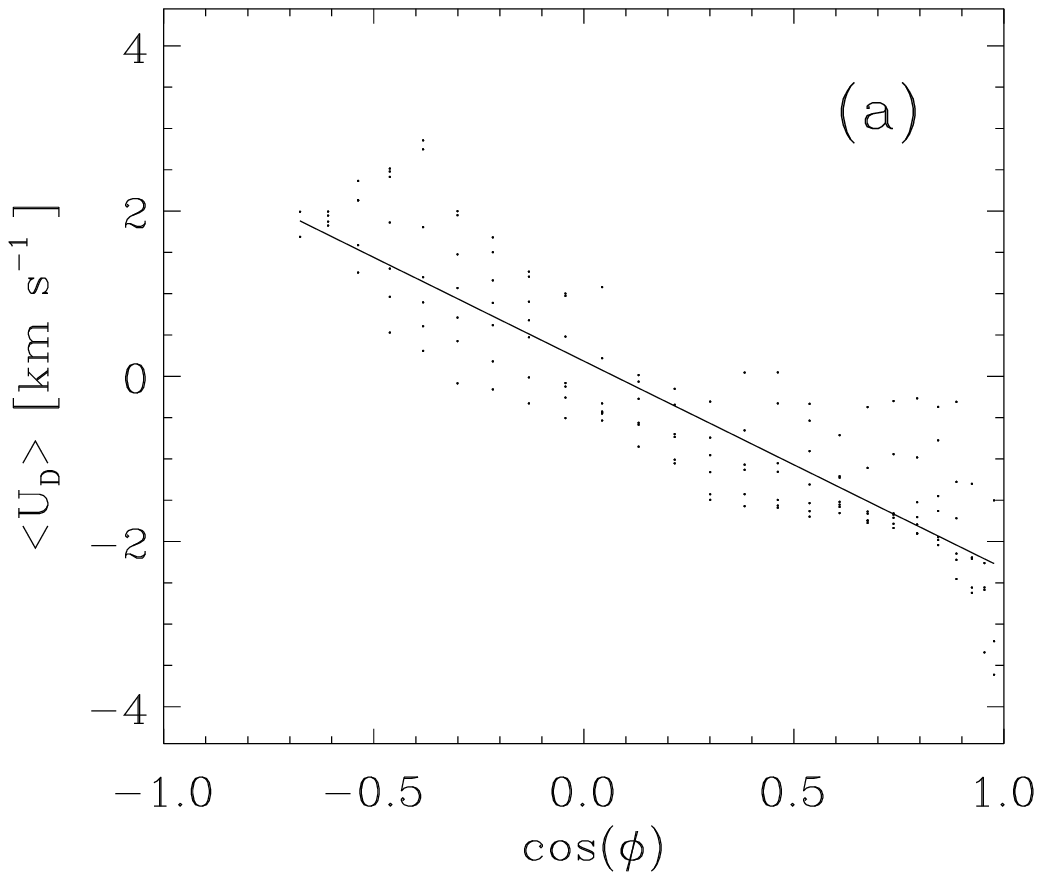}
\plotone{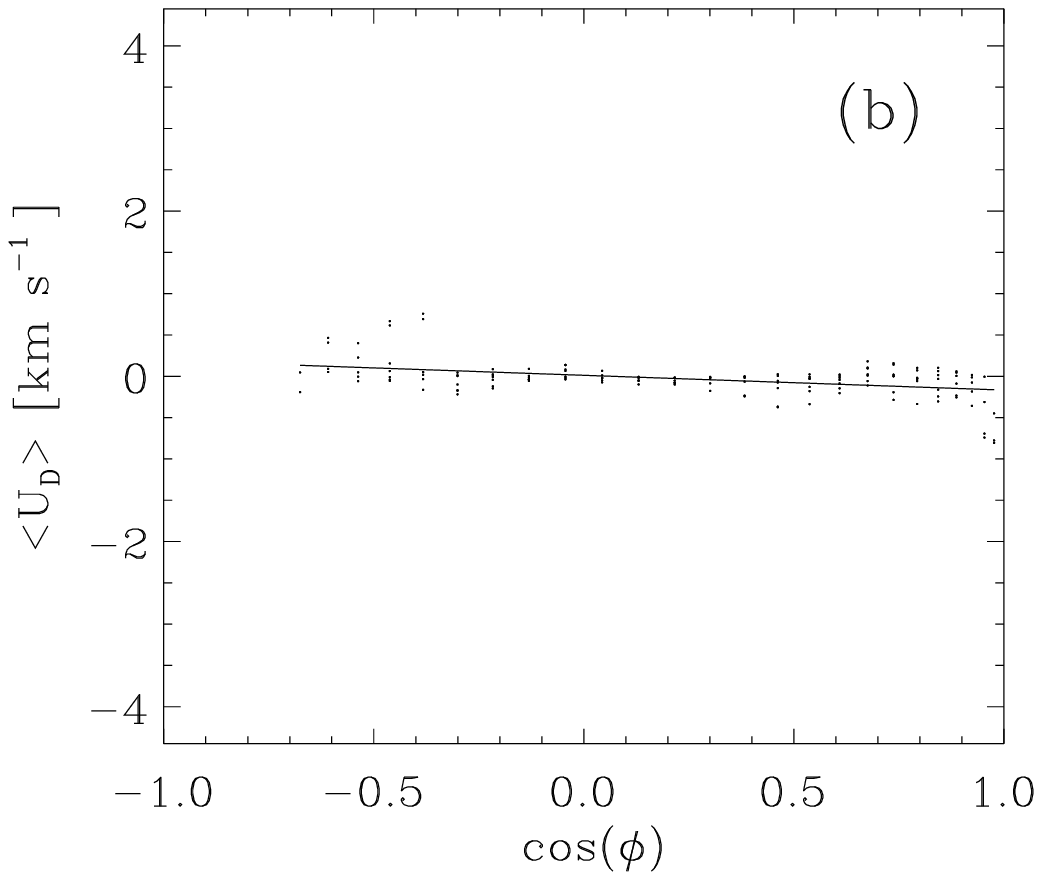}
\caption{Mean Doppler shift versus cosine of the azimuthal
angle for the two Gaussian components that reproduce the 
line asymmetries. (a) Shifted component. It shows
the pattern characteristic of the Evershed flow, 
with blueshifts in the center-side penumbra ($\cos\phi\simeq 1$)
and redshifts in the limb-side penumbra ($\cos\phi\simeq -1$).
(b) Unshifted component, showing very small
Doppler shifts. 
Each point represents the average Doppler shift
over a small range of azimuths and radial distances 
(the sectors represented in Fig.~\ref{maparhophi}). 
Error bars are not included to avoid 
overcrowding, but they are smaller than the scatter 
among individual points.  
}
\label{separates}
\end{figure}

The two components have very different Doppler shifts
(i.e., different $\lambda_i$).
The components in each pixel have been classified 
either as {\em shifted} or as {\em unshifted} 
depending on which one has the largest absolute
Doppler shift. As it is shown in Figure~\ref{separates},
the unshifted component has almost no velocity
throughout the penumbra, whereas the shifted
component follows  the  pattern characteristic of 
the Evershed effect,
with blueshifts in the center-side penumbra and 
redshifts in the limb-side penumbra.  
The velocities of the shifted
component are larger than those inferred from the
bisectors.

\begin{figure}
\plotone{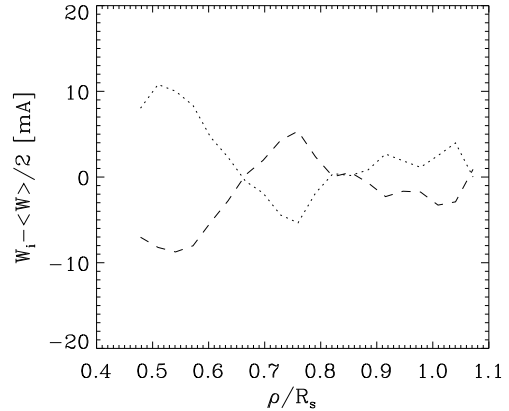}
\caption{Radial variation
of the equivalent widths of the shifted component
(the dashed line) and the unshifted component (the dotted line). 
The same local mean equivalent
width has been subtracted from the two components. 
}
\label{eqwidths}
\end{figure}

The two components contribute similarly to the
line profiles, except in the inner penumbra, where
the unshifted component dominates.
An example of two-Gaussian fit in the
innermost penumbra is shown in Figure~\ref{gauss_fit}, bottom.
Note the broad and shallow shifted component, 
and how the unshifted component dominates the line 
profile. 
Figure~\ref{eqwidths} illustrates the same
result in a more systematic way.
We use the equivalent width to parameterize the
strength of the component.
The equivalent width $W$ is defined
as the area above the line profile
normalized to the continuum intensity. In the case of
Gaussian profiles, 
\begin{equation}
W_i\simeq\sqrt{2\pi}\,d_i\,\sigma_i/S_c(\lambda_c),
\end{equation}
where   we have neglected the variation of the continuum 
within the line profile so that the continuum intensity
$S_c(\lambda)\simeq S_c(\lambda_c)$, with
$\lambda_c$ the line core wavelength.
Figure~\ref{eqwidths} shows the excess of equivalent
width of the two components as a function 
of the distance to the sunspot center.
Following the approach of the previous sections, we
remove a 
running mean 
local average to show 
local variations.
We subtract $\langle W\rangle/2$ with
$\langle W\rangle=\langle W_1\rangle+\langle W_2\rangle$.
This local mean is the same for the two components to
ensure that the differences do not come from
the two components having different local means.
Figure~\ref{eqwidths} shows how the unshifted component 
tends to have  an equivalent width significantly larger than  
the shifted component for $\rho < 0.65\,R_s$, whereas 
the contribution of the two components balances from
this radius on. (The symbol $R_s$ stands for radius
of the sunspot.) Actually, the shifted component
shows a small excess at 0.75, and then the unshifted
component takes over again.  

\begin{figure}
\plotone{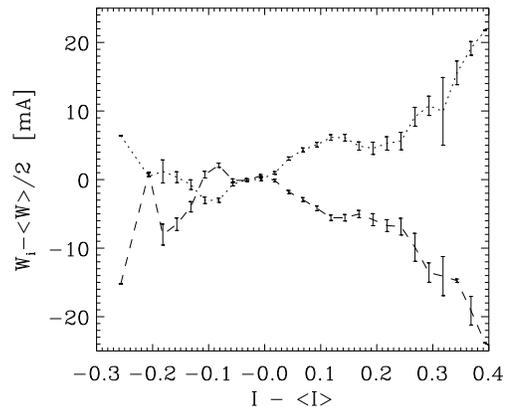}
\caption{Equivalent width of the shifted component (the dashed
line) and the unshifted component (the dotted line).
The unshifted component dominates in locally bright 
structures, suggesting an association between bright features and 
the unshifted component.
The error bars represent the standard deviation of the
averages represented in the figure.
}
\label{equivalent}
\end{figure}

The unshifted component seems to be brighter 
since it dominates in the bright pixels. 
Figure~\ref{equivalent} shows how the equivalent
widths of the two components vary with the local 
intensity. The equivalent width of the unshifted
component is larger in  the bright pixels
($I > \langle I\rangle$),
whereas the contribution of the two components
balance in the locally dark features
($I < \langle I\rangle$). 

\begin{figure}
\plotone{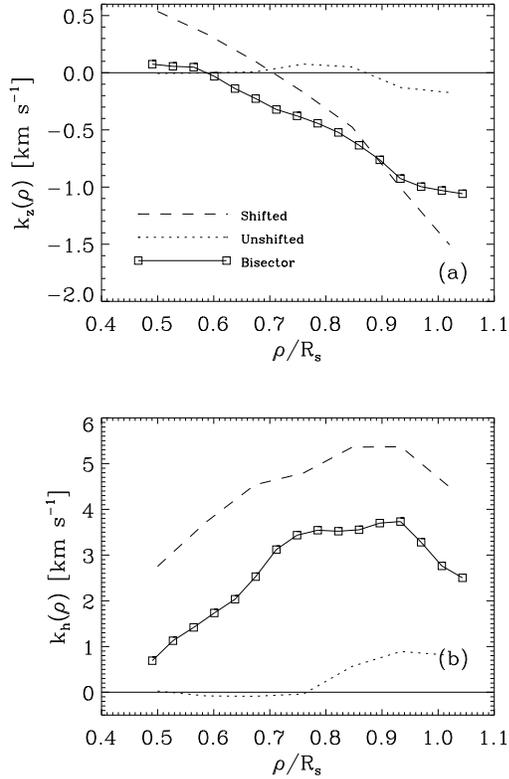}
\caption{
(a) Vertical velocities $k_z$ and (b) horizontal velocities
$k_h$  for the two components resulting from the 
two-Gaussian fits  of the asymmetric line  profiles. We also 
include the velocities inferred from the bisector analysis 
for reference (the solid line with symbols, as in Fig.~\ref{kappas}.) 
The unshifted component presents very small velocities.
Systematic horizontal velocities of up to 
5~km~s$^{-1}$ characterize the shifted component 
(the dashed line).
This component also shows upflows
in the inner penumbra ($k_z >0$ for $\rho/R_s < 0.7$) and
downflows in the external penumbra.
}
\label{kappas_gauss}
\end{figure}

\begin{figure}
\plotone{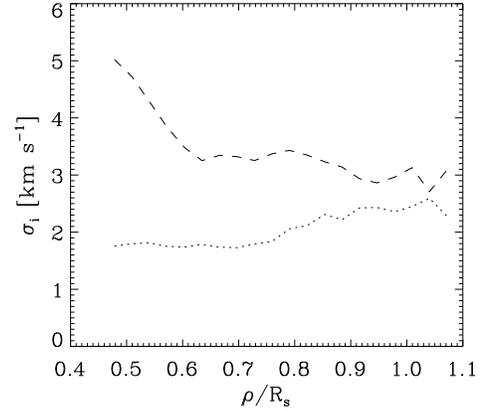}
\caption{
Variation with the radial distance of the widths 
of the two Gaussian components. They are given in 
km~s$^{-1}$ versus radial distance normalized
the Sunspot radius $R_s$. The dotted line corresponds
to the unshifted component whereas the dashed line
represents the shifted component.
(A Gaussian has a FWHM some 2.35 times the parameter
$\sigma_i$ represented in the figure.)
}
\label{widths}
\end{figure}

As we point out above, the Doppler shifts
of the shifted component resembles but exceeds
the Doppler shifts derived from the bisectors.
The corresponding mean vertical and horizontal velocities
can be inferred following the same approach employed for
the velocities based on bisectors. 
Figure~\ref{kappas_gauss} shows $k_z$ and $k_h$ 
obtained from a least squares fit of the Doppler shifts
of the two components using equation~(\ref{avdoppler}). 
Points grouped by radial distance are used
to obtain the radial variation.
(The same approach used to derive Fig.~\ref{kappas}.)
The shifted component shows systematic horizontal velocities
of up to 5~km~s$^{-1}$, and it presents moderate upflows in the
inner penumbra ($k_z < 0.5$~km~s$^{-1}$) and  larger 
downflows in the outer penumbra ($k_z\simeq 1.5$~km~s$^{-1}$).
Simultaneously, the unshifted component shows velocities
never exceeding 1~km~s$^{-1}$.
Because of the quality
of the fits, one can conclude that 
reproducing the observed bisectors requires plasmas
with velocity amplitudes
between 0 and 5~km~s$^{-1}$ 
coexisting in our 0\farcs2 resolution elements.
In the case of the shifted component
the upflows in the inner penumbra exceed
our observational uncertainties ($\sim$100~m~s$^{-1}$;
see \S~\ref{observations}). 
However, we are reluctant to interpret these shifts as 
systematic upflows present in the inner penumbra. If the 
properties of the shifted component are taken
literally, they indicate the existence of spatially
unresolved upflows and downflows of the order of several 
km~s$^{-1}$,
a feature that is more meaningful that 
the small systematic upflow.
The upflow may be real, but it may also be 
a false residual left if the Doppler
signals of the large upflows and downflows are not
strictly proportional to the flow of mass.
The existence of such large upflows and downflows
follows from the analysis of  
Figure~\ref{widths}, which
shows the variation with the radial 
distance of the widths of the two Gaussian components 
($\sigma_1$ and $\sigma_2$ in equation~[\ref{eq2g}]). 
The width of the shifted component reaches a value of 
up to 5 km~s$^{-1}$ in the inner penumbra,
and this value is almost independent of the azimuthal
angle within the sunspot.
This width is too large to be due to thermal motions 
($\sim$ 1 km~s$^{-1}$  for penumbral
temperatures of 5500~K), and it is 
also much larger than the typical widths of the profile of
Fe~{\sc i}~7090.4~\AA\ ($\sim$2 km~s$^{-1}$ in the
quiet Sun). Then if the inferred widths are due to 
spatially unresolved motions, it means that 
redward and blueward velocities  of several km per 
second are associated with the shifted component in the 
inner penumbra. Since the widths do not depend on the
azimuthal angle within the sunspot, such large
velocity dispersion is also representative of the
vertical velocities and, consequently, upflows
and downflows of several km per  second are associated with 
the shifted component. 
According to Figure~\ref{widths}, the width of the shifted 
component suffers a strong decrease from the inner 
to the outer penumbra. 
Observations have consistently shown that the widths
of the spectral lines increase from the 
inner to the outer penumbra 
\citep[e.g.,][]{joh93,rim95,tri04}, 
which seems to be at variance with our result.
However, there is no obvious inconsistency keeping
in mind that the shifted component does not
contribute to the line profile in the inner penumbra
(Fig.~\ref{eqwidths}). The profile is 
dominated by the unshifted component,
which is very narrow and whose
width increases with the radial distance
(Fig.~\ref{widths}). Consequently, the
line width of the full profile is 
expected to go from the values of the
unshifted component in the inner penumbra,
to some kind of average between the widths
of the two components in the outer penumbra,
thus increasing from the inner to the
outer penumbra (see Fig.~\ref{widths}).

A final caveat is in order.
Like the use of bisectors, the two-Gaussian 
fit is only an heuristic representation of the 
observed line asymmetries. 
The two components resulting from our
fits do not necessarily represent two physical
components existing in the penumbra, and
a literal interpretation of the properties of 
the components may be misleading. The Evershed 
effect is not only the result of combining 
two separate components. This description is 
incomplete since two unresolved but symmetric 
components cannot account for the broad-band circular 
polarization observed in sunspots 
(see \S~\ref{discusion}).

\section{Discussion and Conclusions}\label{discusion}

We find local
upflows associated with bright structures and
local downflows associated with dark structures
in 0\farcs2 angular resolution spectra
(\S~\ref{vert_and_hor}).
The correlation keeps the same sign in the limb-side
penumbra and the center-side penumbra, which indicates
that it is due to vertical motions.  
The correlation remains even when only the inner half 
of the penumbra is considered. 
The existence of 
such correlation was originally reported by  \citet{bec69c},
and it is suggestive of convective motions in penumbrae.
It has the  sign
required to transport energy by convection, although
the  magnitude of the inferred velocities 
is not yet enough to account for the observed radiative 
flux (see the discussion below).  
This deficit could be caused by the still 
insufficient spatial resolution of the present data.
Appendix~\ref{appa} shows how the correlation 
in penumbrae is qualitatively similar to the
intensity velocity relationship found in the quiet
Sun when the angular resolution is not enough
to resolve individual convective cells (granules).
In the case of the quiet Sun the correlation 
is produced by convective motions.  The fact that the 
residual correlation is similar to that observed in
penumbrae does not prove the convective nature of the
penumbral pattern. However, the comparison is very 
suggestive.
The relationship between vertical velocity 
and intensity is not the only observable
hinting at convection in penumbrae. For example,
the proper motions arrange the plasma forming long narrow 
filaments  co-spatial with dark features, 
in a behavior characteristic of the convective proper 
motions of the quiet Sun \citep{mar05}. Another hint comes 
from the observed Stokes asymmetries, which can be 
reproduced by small magnetic loops carrying mass flows
that emerge and return to the sub-photosphere 
within the penumbra \citep{san04b}.
If the loops connect a hot footpoint with 
a cold footpoint, then the flow along magnetic field lines 
produces a vertical transport of energy 
\citep[e.g.,][]{sch03f}. 
Such loops may be small scale counterparts
to the filament like structures in velocity maps
found to join bright and dark knots
by \citet{sch05}.

We would like to emphasize that any of the modes of 
convection described in \S~\ref{intro} is in 
principle consistent with the observed correlation
between vertical velocity and intensity. 
The predictions of the various models are simply not 
realistic enough to allow a quantitative comparison with
our observations, which 
should be regarded as a constraint to be 
fulfilled by any further realistic modeling.

According to equations~(\ref{def1}) and (\ref{resul1}),
the relationship between the continuum intensity
fluctuation $\Delta I_{\rm rms}$ and the fluctuations 
of vertical velocity $\Delta U_{z}$ are,
\begin{equation}
\Delta U_{z}
\simeq 3.3~{\rm km\,s}^{-1}
\Delta I_{\rm rms}.
\label{speculate1}
\end{equation}
We observe $\Delta I_{\rm rms}$ to be of the order
of $0.09$,
rendering a $\Delta U_{z}\simeq 0.3$~km~s$^{-1}$
insufficient to account for the radiative
losses of penumbrae. Vertical velocities of the order of 
1~km~s$^{-1}$ are required; see, e.g.,
\citet[][]{spr87}.
However the fluctuations of continuum intensity
are underestimated due to the insufficient
resolution. If the true fluctuations are
as large as $0.2$ rather than the observed value, 
equation~(\ref{speculate1}) yields
$\Delta U_{z}\simeq$ 0.7~km~s$^{-1}$, 
which  could easily account for the radiative 
losses of penumbrae. Such large continuum 
contrast is characteristic of numerical 
simulations of non-magnetic convection in the quiet Sun
(e.g., \citeauthor{stei98}~\citeyear{stei98}
obtain $\Delta I_{\rm rms}\sim 0.16$).
The above estimate implicitly assumes 
the solar surface to be fully covered by 
upflows and downflows. This simplifying 
hypothesis is not needed. Following arguments 
similar to  those by \citet{spr87}, \citet{sch03f} 
estimate the energy transported by the moving 
fluxtube model of interchange convection 
\citep{sch98a}. The model fluxtubes have very fast 
hot stationary upflows which are shown to be
capable of explaining the penumbral
radiative losses even if the fluxtubes fill only a 
small fraction of the  surface (a coverage of some 
10\% suffices for upflows of 4 km~s$^{-1}$). 
The estimate by \citet{sch03f} 
is also valid in a general case 
of concentrated upflows and downflows.
They can be consistent with the radiative 
losses of penumbra and with our observation,
provided that the pattern of concentrated 
upflows and  downflows,
once smeared to our spatial resolution, 
gives rise to the intensity vertical
velocity correlation described 
in \S~\ref{vert_and_hor}.

The velocities discussed above are computed using the
shift of the bisector in the wings of the 
non-magnetic line Fe~{\sc i}~7090.4~\AA\ .
The use of bisectors closer to the line core does 
not modify the scenario described above, although 
the velocities are smaller. However, the fact that the 
bisector shift depends on the intensity level within the
line is consequential. It proves in a direct and  
robust manner that the penumbral 
velocity structure remains unresolved in 
our spectra. 
This result is not unexpected, though. 
Images of penumbra show structures at least a factor 
two thinner than our resolution \citep{sch02,rou04},
and it is conceivable that part of such structuring
is shared by the velocity pattern. 
Even more direct evidence for unresolved velocities
comes from the presence of broad-band circular 
polarization (BBCP) in penumbrae, discovered thirty years
ago by \citet{ill74a,ill74b}.
The BBCP demands large gradients of velocity along the 
line-of-sight within the range of heights where a typical 
photospheric line is formed \citep[e.g., ][]{san92b,sol93b}.
This range is not larger than 150~km and therefore unless
the spatial resolution is significantly better than
this value, spectral lines with slanted bisectors 
are to be expected.

Even our best spatial 
resolution does not allow us to resolve
the Evershed velocity pattern, which rises
an obvious concern. All analyses of 
penumbral spectra assuming a single resolved 
component 
may lead to biased velocity estimates.
The existence of a single
component is an implicit assumption of all
those measurements that assign a single 
velocity to each resolution element. Often
this simplifying assumption allows 
to observe with a spatial resolution 
otherwise impossible to reach 
\citep[e.g.][]{lan05}.
However, these measurements are liable to a
significant bias
since reproducing 
the observed  line shapes requires plasmas with 
velocities between 
0 and 5~km~s$^{-1}$  
co-existing in the resolution elements. 
This range of velocities is deduced from the two 
component fits carried out in \S~\ref{two_gauss_fit}, 
but it is also very consistent with  the dispersion of 
velocities inferred from the asymmetries of the 
Stokes profiles \citep[e.g.,][]{wes01b,mat03,san04b}.
If the properties of the observed spectra are 
measured assuming a single component per resolution
element, the measurement only provides an ill-defined 
average of the unresolved velocities.
Discrepancies between the measured velocities
and the true mass weighted average velocities are to be 
expected, with differences that can be a fraction 
of the dispersion of velocities in the resolution 
element. Since the dispersion  of velocities in the 
penumbra is of the order of a few km~s$^{-1}$,
differences as large as one km~s$^{-1}$ are 
conceivable.
Obviously, this caveat also affects the velocities 
measured when working out the intensity velocity 
relationship  in \S~\ref{vert_and_hor}.
Note, however, that these are differential 
measurements unaffected by a global bias. 
Only differences of the bias between bright and dark 
features are of concern, and they are expected to be 
of second order.

In addition to the correlation between intensity and 
vertical velocity  pointed out  
above, we find two other 
results in  good agreement with previous observations.
First, the {\em horizontal} velocity increases 
in dark penumbral features. This conclusion follows 
from the negative slope of the correlation between 
horizontal motions and intensities ($m_h < 0$ in 
equation~[\ref{resul1}]).
Second, we find moderate but systematic downflows in
the penumbra\footnote{As we argue in
the previous paragraph, this may not represent
a net vertical motion of the plasma but a bias produced
by the imperfect cancellation of unresolved 
upflows and downflows.}. The presence
of such downflows over a significant part of the penumbra
is a result consistently shown by most recent measurements  
(see Fig.~\ref{kappas}a and the references in there).
We want to emphasize these two non-trivial 
coincidences with previous works before discussing 
apparent disagreements with two other recent 
observational papers. Such disagreement is 
particularly discomforting since the works 
also employ high spatial resolution
SST data.
Because of 
the potential importance, we discuss the
disagreement  in some detail.
\citet{lan05} do not find the systematic downflows 
in penumbrae discussed above. Their measurements
are optimized for magnetic studies, therefore, they
employ a magnetic line whose changes of shape 
due to magnetic field variations may induce 
spurious velocity signals. They use filtergrams 
with moderate spectral resolution (72~m\AA\ FWHM),
thus averaging information coming from different 
line depths. Only two wavelengths within the 
line profile are sampled, and these are not 
far from the line core ($\pm 50$~m\AA ). 
These parameters suggest a weighting of their measurements 
towards the line core, where the 
systematic  downflows that we find are milder
(the bisector at the 20\%  intensity level has 
downflows only 1/3 of the values in Fig.~\ref{kappas}a).  
All these factors combined may eventually 
explain the lack of systematic downflows.
\citet{bel05} find dark cores
in penumbral filaments having upflows instead of the
downflows to be expected if a dark
structure follows the intensity vertical velocity 
relationship in \S~\ref{vert_and_hor}. 
\citet{bel05} employ the same instrumentation used in
our work. Even the same spectral line is used 
in some of the estimates. Consequently, the issue of 
spectral resolution argued above is of no 
relevance for sorting out discrepancies. 
The differences must be pin down to the technique 
of analysis. One can identify two main 
differences with respect to our case.
First, they use as absolute velocity reference
the wavelength of the core of the spectral line
in the dark cores. According to \citet{bel05}, this 
reference provides an absolute reference within a 
"few hundred m~s$^{-1}$". However,
in the inner penumbra, where the dark cores
reside, a shift of the absolute velocity scale by
200 m~s$^{-1}$ suffices to turn upflows into 
downflows and vice-versa  (see Fig.~\ref{kappas}a 
for $\rho/R_s<$0.7).   
Second, they study a limited number of specific dark 
cores. The correlation that we find is not one-to-one
and those cases may correspond to deviations
from the mean law. In this sense, the dark 
features giving rise to the correlation between intensity
and velocity may not correspond to dark cores but a
different penumbral dark structure. In short,  the
reasons for the apparent discrepancies between our 
results and these two SST based works are so 
far unclear and must be investigated
with new observations.
They will also allows to correct an obvious limitation 
of our work. It is based on one half of the 
penumbra of a sunspot observed in a single 
position on the disk. The used of a data set 
with such poor statistics is justified by the 
exceptional angular resolution of the spectra,
and by the fact that Evershed flow seems to 
be a property common to all penumbra. However,
the results require confirmation.

\acknowledgements

Thanks are due to Dan Kiselman and the SST staff for support
during the observations.
The SST is operated by the Institute
for Solar Physics, Stockholm, at the Observatorio
del Roque de los Muchachos of the Instituto  de Astrof\'isica
de Canarias (La Palma, Spain). 
The work has partly been funded by the Spanish Ministry 
of Science and Technology, project AYA2004-05792.
A preliminary analysis of this data set was 
presented to the {\em Solar Polarization Workshop 4}, 
and a summary will appear in the 
proceedings \citep{bon06}.
%
\appendix

\section{Effect of stray-light on the absolute 
	wavelength scale}\label{appb}

The spatial stray-light contaminates the average umbral 
spectrum used to set the absolute wavelength scale. 
The observed umbral spectrum $S(\lambda)$ is
not the true umbral spectrum $S_u(\lambda)$ but
it contains a fraction $\alpha$ of stray-light spectrum 
$S_s(\lambda)$,
\begin{equation}
S(\lambda)=\alpha\,S_s(\lambda)+(1-\alpha)\,S_u(\lambda).
\label{appb_eq1}
\end{equation} 
The contamination artificially shifts $S(\lambda)$
with respect to $S_u(\lambda)$ if  $S_s(\lambda)$
is shifted with respect to  $S_u(\lambda)$
by an amount $\lambda_s\not= 0$. For the sake
of simplicity, we assume the stray-light spectrum 
to have a shape similar to the umbral spectrum, 
therefore,
\begin{equation}
S_s(\lambda)\simeq C^{-1}\, S_u(\lambda-\lambda_s),
	\label{maria0}
\end{equation} 
with the constant $C$ standing for the ratio 
between the continuum intensities of the umbra and 
the stray-light. To first order in $\lambda_s$,
\begin{equation}
S_s(\lambda)\simeq C^{-1}\,\Big[S_u(\lambda)-\lambda_s
	{{dS_u(\lambda)}\over{d\lambda}}\Big],
\end{equation} 
and inserting the previous expression into 
equation~(\ref{appb_eq1}),
\begin{equation}
S(\lambda)\simeq (1-\alpha+\alpha/C)\,S_u(\lambda-\lambda_\alpha),
	\label{maria1}
\end{equation} 
\begin{equation}
\lambda_\alpha=
\lambda_s\Big/\Big[1+{{1-\alpha}\over{\alpha}}C \Big],
\label{appb_eq0}
\end{equation}
where $\lambda_\alpha$
gives to the artificial shift introduced 
by the stray-light. 
%
In order to estimate $\lambda_\alpha$, one needs to know
the fraction of stray-light $\alpha$, the 
shift of the stray-light contamination $\lambda_s$, and
the continuum contrast of the stray-light $C$.
We consider 
\begin{equation}
\alpha\leq 0.05,
\label{alpha}
\end{equation}
to be a realistic upper limit to the stray-light 
contamination. The brightness of  the
Venus disk in SST images taken during the 
June 2004 Venus 
transit\footnote{http://www.solarphysics.kva.se/.} 
provides the level of SST stray-light. At the 
wavelength of our observation, the Venus disk
shows an intensity smaller than 0.05 times
the quiet Sun intensity as soon as it is measured  
1\arcsec\ inside the Venus limb. The umbral 
stray-light  signals come from the penumbra
or beyond.  
The umbral region used for wavelength 
calibration is separated from the penumbral border by more 
than 1\arcsec, which justifies the upper limit 
in equation~(\ref{alpha}).
The shift of the stray-light profile is 
unknown but, if the stray-light signals come 
from the neighboring penumbra, it can be estimated 
from the line core shift of the observed 
penumbral profiles. The shift of the mean penumbral 
profile turns out to be of only 30~m~s$^{-1}$.
The shift considering only the inner penumbra 
is even smaller. These figures are effected 
by the uncertainty that we are trying to estimate
and, therefore, they must be taken with some 
caution. However, the mean penumbral profile making up
the stray-light contamination is expected
to present a shift much smaller than the 
large Evershed shifts discussed in the paper. First, 
we are using the line core for wavelength
calibration, where the Evershed shifts
are largely reduced. Second, the Evershed shifts of the 
center-side penumbra and limb-side penumbra have 
opposite signs so that they tend to cancel in the 
averages.  Only the (small) vertical component 
of the Evershed effect is left. 
We use as an upper limit for 
$\lambda_s$ the convective blueshift of 
Fe~{\sc i}~7090.4~\AA\ in the quiet Sun,
\begin{equation}
|\lambda_s|\leq 300~{\rm m~s}^{-1};
\label{beta}
\end{equation}
see \citet[][ Fig. 6b]{dra81}.
By considering this value, we are also including
the case where the stray light
is not produced in the penumbra but further out.
Finally,
\begin{equation}
C\simeq 0.3,
\label{ccc}
\end{equation}
a figure coming from the continuum 
intensity in our umbra when  
$\alpha=0.05$, and when the stray-light 
intensity is given by the quiet Sun intensity.
(We use equations~[\ref{maria0}] and [\ref{maria1}]
at continuum wavelengths to evaluate $C$.)
Equations~(\ref{appb_eq0}) (\ref{alpha}) 
(\ref{beta}) and (\ref{ccc}) yield,
\begin{equation}
|\lambda_\alpha|\leq 45~{\rm m\,s}^{-1},
\label{appb_eq2}
\end{equation}
which sets an upper limit to the effect
of the stray-light on our absolute
wavelength scale.
A final comment is in order. 
We do not consider the {\em spatial} stray-light 
produced by the spectrograph because it is believed
to be negligible. The slit of the spectrograph
selects only a very small portion of the solar
surface.

\section{Intensity-velocity correlation in low resolution granulation
images}\label{appa}

\begin{figure}
\plottwo{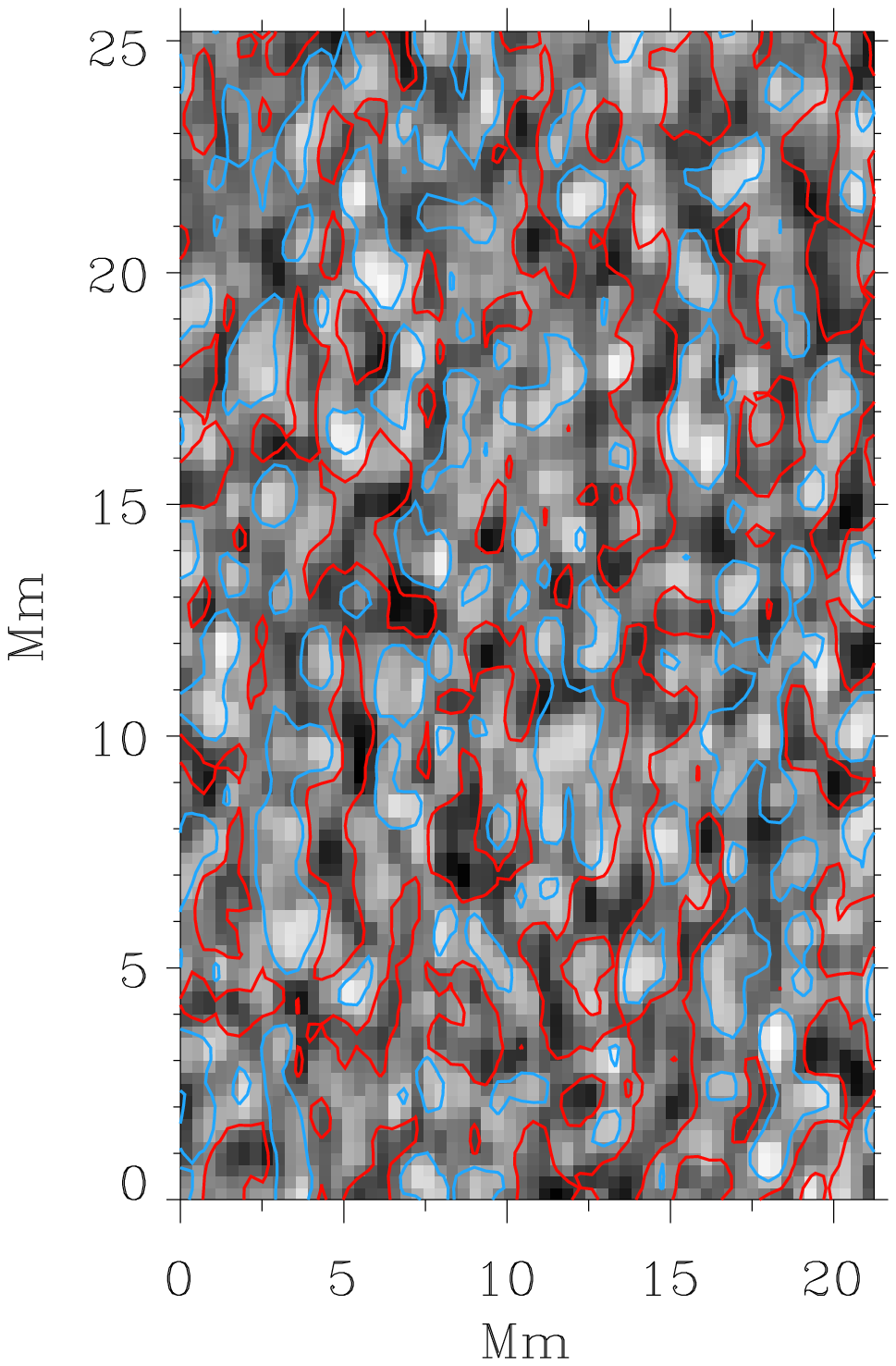}{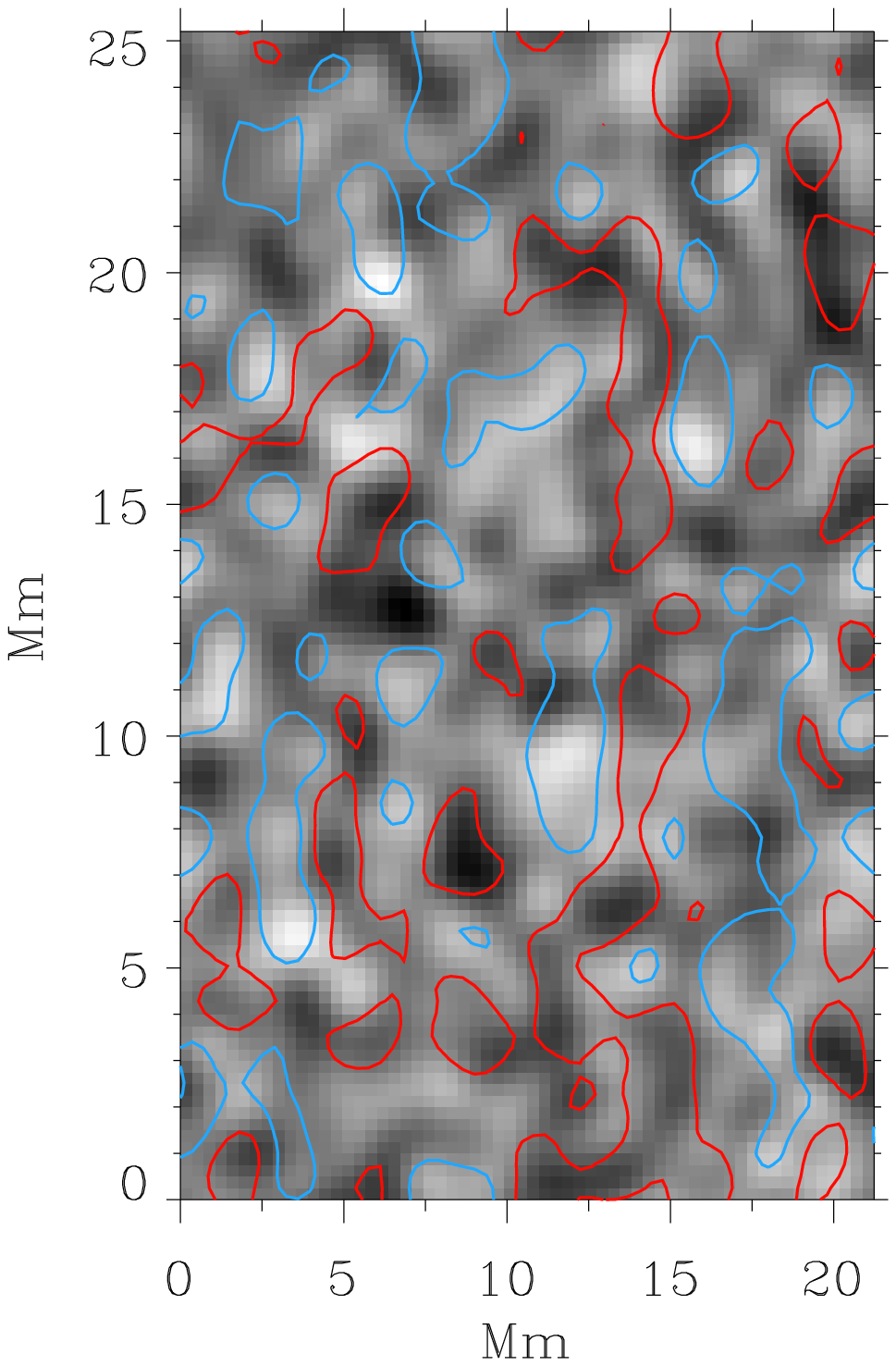}
\caption{Continuum intensity and Doppler velocity contours
in a quiet Sun region at the solar disk center. 
Left: original 1\arcsec\ resolution observations. The
contours correspond to $\pm$250 m~s$^{-1}$.
Right: same observation smeared with a  2\arcsec\ 
Gaussian, and with velocity
contours at $\pm$150 m~s$^{-1}$
Note the correlation between upflows 
(blue contours) and bright features. The correlation 
is not perfect, decreasing as the resolution worsens.
}
\label{ita_quiet}
\end{figure}

The observed correlation between intensity
and Doppler shift is not one-to-one. As we mention in the main
text, a clear correlation can be  
washed out due to the still insufficient resolution of 
the observations, which may not allow us to see 
individual penumbral convective cells. In order to illustrate 
the effect, we have carried out an analysis similar to that 
yielding Figure~\ref{ita_like} but using low spatial
resolution quiet Sun granulation. Figure~\ref{ita_quiet}a
shows intensities and velocities obtained from the disk center 
1\arcsec\ resolution Fe {\sc i}~15648~\AA\ observations 
described by \citet{san03c} and  \citet{dom06b}.  
(These spectra were used for convenience,
but the behavior should be representative of any 
other line.)
The Doppler shifts have been measured as the barycenter of the
spectral line, which is fairly symmetric.
One can readily see the correlation between bright features
and upflows, the latter shown as blue contours in the figure.
Note how the correlation is not perfect, so that
sometimes redshift contours overlay locally bright features,
and vice-versa.
Figure~\ref{ita_quiet}b shows the maps in Figure~\ref{ita_quiet}a
smeared with a Gaussian 2\arcsec\ wide. The correlation is
still visible but it worsens. The 2\arcsec\ resolution
maps could be comparable to our penumbral data assuming the penumbral convective
cells to be 0\farcs 1 wide or half our spatial resolution
(\S~\ref{observations}). The granulation cells are some 1\arcsec\ wide and
so half the resolution of  Figure~\ref{ita_quiet}b.
This simple numerical experiment shows how
the quiet Sun convection observed with insufficient
angular resolution produces maps of intensity and velocity
showing only a moderate correlation, despite the fact
that the intrinsic local 
correlation must be very high \citep[see, e.g., Fig.~3 of the
realistic simulations by ][]{stei98}.

%


\end{document}